\documentclass[a4paper,fleqn,usenatbib]{mnras}
\usepackage[T1]{fontenc}
\usepackage{ae,aecompl}

\usepackage{graphicx}	
\usepackage{amssymb}	
\usepackage{mathrsfs}

\usepackage[usenames]{color}
\usepackage{amsmath}

\newcommand {\bc}{\begin{center}}
\newcommand {\ec}{\end{center}}
\newcommand {\be}{\begin{equation}}
\newcommand {\ee}{\end{equation}}
\newcommand {\beq}{\begin{eqnarray}}
\newcommand {\eeq}{\end{eqnarray}}

\newcommand {\comment}[1]{}

\renewcommand{\d}{{\rm d}}

\title[Propagating mass accretion rate fluctuations]
{Propagating mass accretion rate fluctuations in X-ray binaries under the influence of viscous diffusion}
\author[A. A.~Mushtukov et al.] 
{Alexander~A.~Mushtukov,\thanks{E-mail: al.mushtukov@gmail.com (AAM)}  
Adam~Ingram, and
Michiel~van der~Klis \\ 
 Anton Pannekoek Institute, University of Amsterdam, Science Park 904, 1098 XH Amsterdam, The Netherlands \\
} 
\date{Accepted 2017 September ??. Received 2017 September ??; in original form 2017 July ??}

\pubyear{2017}

\begin{document}
\label{firstpage}
\pagerange{\pageref{firstpage}--\pageref{lastpage}}

\maketitle

\begin{abstract}
Many statistical properties of X-ray aperiodic variability from accreting compact objects can be explained by the propagating fluctuations model applied to the accretion disc. The mass accretion rate fluctuations originate from variability of viscosity, which arises at every radius and causes local fluctuations of the density. The fluctuations diffuse through the disc and result in local variability of the mass accretion rate, which modulates the X-ray flux from the inner disc in the case of black holes, or from the surface in the case of neutron stars. A key role in the theoretical explanation of fast variability belongs to the description of the diffusion process. The propagation and evolution of the fluctuations is described by the diffusion equation, which can be solved by the method of Green functions. We implement Green functions in order to accurately describe the propagation of fluctuations in the disc. 
For the first time we consider both forward and backward propagation.
We show that 
(i) viscous diffusion efficiently suppress variability at time scales shorter than the viscous time, 
(ii) local fluctuations of viscosity affect the mass accretion rate variability both in the inner and the outer parts of accretion disc, 
(iii) propagating fluctuations give rise not only to hard time lags as previously shown, but also produce soft lags at high frequency similar to those routinely attributed to reprocessing, 
(iv) deviation from the linear rms-flux relation is predicted for the case of very large initial perturbations. 
Our model naturally predicts bumpy power spectra.
\end{abstract}


\section{Introduction}
\label{intro}

Accreting black holes (BH) in active galactic nuclei (AGN) and BH binaries as well as neutron star (NS) binaries show strong variability of the X-ray flux over a very broad (several decades) frequency range: the time scale extends down to milliseconds in binary systems and minutes in AGN \citep{2000A&A...363.1013R,2004MNRAS.348..783M}. 
The broad band component of the power density spectrum (PDS) is generally well described by a twice-broken power-law. In the case of some BH binaries, the PDS displays a narrow feature peaking in the range $\sim\,0.1-10\,{\rm Hz}$s indicated as quasi-periodic oscillations (QPOs). Light curves both in BH and NS systems show a log-normal flux distribution and a linear relationship between the root-mean square (rms) variability and the luminosity of the source \citep{2001MNRAS.323L..26U,2002PASJ...54L..69N,2004MNRAS.347L..61U,2004ApJ...612L..21G,2005MNRAS.359..345U}. 

The light curves in different energy bands are correlated with each other, but the hard X-ray variability usually lags the soft X-ray variability \citep{1979ApJ...233..350P,1981ApJ...246..494N}. The magnitude of the time-lag depends on frequency, but typically it is of order of 1 per cent of the variability time scale. 
Some sources show negative time-lags, with the soft-band variability lagging the hard-band variability \citep{2007MNRAS.382..985M,2011MNRAS.416L..94E,2013MNRAS.431.2441D}. This type of behavior has been detected in both supermassive and stellar mass BH.
The cross-correlation function between any two energy bands peaks at zero lag \citep{2000ApJ...537L.107M,2001AIPC..599..310P}, 
and emission from different energy bands is coherent at low frequencies and incoherent at relatively high frequencies \citep{1999ApJ...510..874N}. It is interesting that the same features are observed in the case of accreting non-magnetized white dwarfs \citep{2013MNRAS.431.2535S} and even young stellar objects \citep{2016MNRAS.463.2265S}. 

The observed variability is naturally explained by the propagating fluctuations model \citep{1997MNRAS.292..679L}, in which  fluctuations arise at different radial coordinates of the accretion disc and then propagate towards the central object. Different time scales are injected into the accretion flow at different distances from the central object  \citep{1997MNRAS.292..679L,2001MNRAS.321..759C,2015ebha.confE...8I}. More specifically, the initial perturbations of viscosity lead to perturbations of the local mass accretion rate and surface density. The perturbations propagate through the accretion disc and cause local variability of the mass accretion rate at all radii. The variability of the local mass accretion rate leads to photon flux variability observed from different parts of the accretion disc. The propagating fluctuations model naturally explains the lagging of hard X-ray variability \citep{2001MNRAS.327..799K,2013MNRAS.434.1476I}.
Simplified Green functions were used for propagating fluctuations modeling of aperiodic X-ray variability by \cite{1997MNRAS.292..679L} and \cite{2001MNRAS.327..799K}, where the fluctuations were assumed to be additive. Multiplicative fluctuations in the accretion disc were considered by \cite{2013MNRAS.434.1476I} and \cite{2011MNRAS.415.2323I} but in all cases only inward propagation was considered.

In the case of accretion onto magnetized NSs the accretion disc is disrupted at the magnetospheric boundary, which is much larger than the NS radius. From the magnetospheric radius, matter follows the magnetic field lines and reaches the NS surface in small regions located at the magnetic poles, where most of the kinetic energy is released in the form of X-rays. The mass accretion rate variability produced in the accretion disc results in mass accretion rate variability at the magnetospheric radius producing variability of the final X-ray flux \citep{2009A&A...507.1211R}. Thus, the basic physical processes of generating mass accretion rate fluctuations and their evolution in the accretion disc that can be diagnosed by the X-ray flux variability are similar for BH and NS X-ray binaries (XRB). 

Accretion onto weakly magnetized NSs results in a boundary layer between the accretion disc and NS surface, where a significant fraction of total X-ray flux is produced \citep{2010AstL...36..848I}. The variability in this case is strongly affected by processes within the boundary layer, which is beyond the scope of this paper.

{The propagating mass accretion rate fluctuations are determined by the geometry of the accretion disc and the physical conditions in the disc (in particular local viscosity and geometrical thickness). On the other hand, the propagating fluctuations can influence the general features of accretion disc such us, under certain conditions, its stability (see e.g. \citealt{2012A&A...540A.114J}).}

In order to explain the fast variability of accreting BHs and NSs it is necessary to precisely describe the processes of generation and subsequent propagation of fluctuations in the accretion disc. In this paper we focus on the propagation of fluctuations by viscous diffusion. The propagation of mass accretion rate fluctuations can be described by the diffusion equation written under specific assumptions about the accretion disc structure. In some cases the solution of the diffusion equation can be simplified and obtained in the form of Green functions \citep{1974MNRAS.168..603L}. In general the diffusion equation has to be solved numerically  \citep{2015arXiv151102356M}. 

The key principles of the construction of a Green function for the diffusion equation of disc accretion was formulated by \cite{1952Lust}. The specific Green functions obtained so far in the literature were obtained for the case of a Newtonian potential and various sets of initial assumptions encoded in boundary conditions. (i) The Green function for the case of an infinite accretion disc with the inner radius located at $R_{\rm in}=0$ was derived by \cite{1974MNRAS.168..603L}. \cite{1991MNRAS.248..754P} discussed disks with a central source of angular momentum. (ii) The Green functions for the case of nonzero inner radius but infinite outer radius $R_{\rm out}$ were found by \cite{2011MNRAS.410.1007T}. This particular Green function can be useful for the description of accretion discs around magnetized sources, where the inner disc radius is defined to be the size of magnetosphere. (iii) The evolution of accretion discs with fixed outer radius was discussed by \cite{1998MNRAS.293L..42K} and \cite{2015ApJ...804...87L}. The same problem was solved numerically by \cite{2009MNRAS.399.1633Z} and a particular case was discussed by \cite{2001ApJ...563..246W}.

In this paper we discuss how Green functions can be included in the model in order to describe propagation of fluctuations of mass accretion rate through the accretion disc, {which is assumed to be radiatively effective (i.e. most of the viscously dissipated energy is radiated locally)}. 
{For the first time we consider the exact description of the diffusion process in application to propagating mass accretion rate fluctuations. This description requires propagation of the fluctuations not only inwards but also outwards, which has not been considered before.}
We describe how the resulting aperiodic X-ray flux variability in different energy bands is affected by diffusion processes. 
{We reproduce the basic features of propagating mass accretion rate fluctuations discussed earlier in literature and get the basic relations as a limiting cases of our general approach. At the same time we show that the accurate description of the diffusion  process implies features that have not been expected before: we show that high frequency mass accretion rate variability at larger radii lags the variability at smaller radii under certain conditions. It leads to important conclusions on the nature of soft lags observed in supermassive and stellar mass BHs.}
{For the first time} we consider the possibility of interactions between fluctuations. Our numerical results use the Linden-Bell/Pringle Green function \citep{1974MNRAS.168..603L}, but the developed formalism allows the use of any appropriate Green function. Though the Green functions we use are constructed for flat space time, it is possible to get basic features arising from a detailed description of propagation fluctuation process. The results are mainly addressed to BH XRBs, but the technique we apply is applicable to accreting magnetized NS binaries.

\section{Viscous diffusion in an accretion disc}

\subsection{The equation of viscous diffusion}

We consider a thin Keplerian accretion disc with geometrical thickness $H$ much smaller than the radial scale $R$: $H\ll R$. The gravitational potential is given by
\be\label{eq:NewtonianPot}
\phi_{\rm N}=-\frac{GM}{R}, 
\ee
where $M$ is a mass of a central object. The Keplerian angular velocity is $\Omega_{\rm K}=\sqrt{GM/R^3}$. Differential rotation and viscous stress lead to an angular momentum exchange between adjacent rings. As a result, angular momentum is transported outwards.

The viscous force is dissipative and the work it does on adjacent rings goes into heat.
The viscous stress $t_{r\varphi}$ is defined as the viscous force per unit area. It is proportional to the shear and can be written as $t_{r\varphi}=\eta R \d\Omega_{\rm K}/\d R$, where the coefficient of proportionality $\eta$ is called dynamic viscosity. The kinematic viscosity is defines as $\nu=\eta/\rho$, where $\rho$ is local mass density.

The evolution of the disc surface density due to viscous processes is described by the diffusion equation, which can be obtained from the laws of conservation of energy and momentum (e.g. \citealt{2002apa..book.....F}):
\be\label{eq:DifEqGen}
\frac{\partial \Sigma(R,t)}{\partial t}=\frac{1}{R}\frac{\partial}{\partial R}\left[R^{1/2}\frac{\partial}{\partial R}\left(3\nu \Sigma R^{1/2}\right)\right],
\ee
where $\Sigma$ is the surface density and $t$ is time. The kinematic viscosity $\nu$ has the meaning of a diffusion coefficient. 
It has to be mentioned that the evolution of the surface density is expected to be described by diffusion process on sufficiently long time scales only. On very short time scales the variability is defined by other processes including propagation of sound and magnetohydrodynamic (MHD) waves.

According to the $\alpha$-prescription \citep{1972AZh....49..921S,1973A&A....24..337S} the viscous stress is described by $t_{r\varphi}=\alpha P$, {where $P$ is the gas pressure in the disc and $\alpha<1$ is a numerical factor.}
\footnote{{The total pressure in the accretion disc is composed of gas pressure and radiation pressure. If the viscous stress is parameterised by total pressure dominated by radiation pressure, the disc is locally unstable \citep{1974ApJ...187L...1L}. However, if the stress is due to magnetic turbulence it scales with gas pressure alone \citep{1989ApJ...342...49S}.}}
It is equivalent to the prescription 
\be\label{eq:AlphaPrescrNu}
\nu=\frac{2}{3}\alpha c_{\rm s}H=\frac{2}{3}\frac{\alpha c^2_{\rm s}}{\Omega_{\rm K}},
\ee
where $c_{\rm s}=(P/\rho)^{1/2}$ is the local isothermal sound speed.

The diffusion equation (\ref{eq:DifEqGen}) can be simplified by using specific angular momentum $h=(GMR)^{1/2}$ as a variable instead of $R$. Then the equation is reduced to \citep{1974MNRAS.168..603L}
\be\label{eq:DifEqN}
\frac{\partial \Sigma}{\partial t}=D_{\rm N}(h,M)\frac{\partial^2 \left(h\nu\Sigma\right)}{\partial h^2},\quad
D_{\rm N}(h,M)=\frac{3}{4}\frac{(GM)^2}{h^3}.
\ee

\subsection{Green function}

In general the diffusion equation is not linear because the kinematic viscosity $\nu$ can be a function of local physical conditions (for example, surface density or local temperature). But if this is not the case and the kinematic viscosity is a function of radius only, the diffusion equations (\ref{eq:DifEqGen}) and (\ref{eq:DifEqN}) are linear and can be solved by using the Green function method \citep{1952Lust}. Then the solution of the diffusion equation is given by
\be
\Sigma(R,t)=\int\limits_{R_{\rm in}}^{R_{\rm out}}G(R,R',t-t_0)\Sigma(R',t_0)\d R' ,
\ee
where $G(R,R',t)$ is a Green function, $\Sigma(R,t_0)$ is the surface density distribution over the radial coordinate $R$ at $t=t_0$, and $R_{\rm in}$ and $R_{\rm out}$ are inner and outer disc radius respectively. The Green function is defined by the kinematic viscosity $\nu(R)$, the inner and outer disc radii and the boundary conditions there.

The local mass accretion rate is defined by the surface density and radial velocity of accretion 
$v_{\rm R}$: $\dot{M}(R,t)=2\pi R\Sigma(R,t)v_{\rm R}$. It can be also expressed as a function of surface density and viscosity:
\beq\label{eq:mdot00}
\dot{M}(R,t)=6\pi R^{1/2}\frac{\partial}{\partial R}\left(\nu\Sigma(R,t)R^{1/2}\right).
\eeq
If the kinematic viscosity is given by a power law $\nu\propto R^n$ the mass accretion rate $\dot{M}(R,t)\propto R^{1/2} \frac{\partial}{\partial R}\left(\Sigma(R,t)R^{n+1/2}\right)$ and can be represented by Green functions as well:
\beq\label{eq:mdot01}
\dot{M}(R,t)=\int\limits_{R_{\rm in}}^{R_{\rm out}}G_{\dot{M}}(R,R',t-t_0)\Sigma(R',t_0)\d R', 
\eeq
where $G_{\dot{M}}(R,R',t)$ is a Green function for the mass accretion rate. 
If the initial surface density fluctuations on top of the diffusion process are described by $\Sigma^*(R,t)$, then the resulting fluctuations of the mass accretion rate are given by a convolution \citep{1997MNRAS.292..679L}: 
\beq\label{eq:mdot01_}
\dot{M}(R,t)=\int\limits_{R_{\rm in}}^{R_{\rm out}}G_{\dot{M}}(R,R',t)\otimes_t \frac{\partial \Sigma^*(R',t)}{\partial t}\d R' 
\eeq
($\otimes_x$ is sing of convolution in $x$-variable).
According to (\ref{eq:mdot00}) the Green function for the mass accretion rate $G_{\dot{M}}(R,R',t)$ is
\be 
\label{eq:mdot02}
G_{\dot{M}}(R,R',t)=6\pi R^{1/2}\frac{\partial}{\partial R}\left(\nu G(R,R',t)R^{1/2}\right).
\ee

If the kinematic viscosity does not depend on time and is given by 
\be\label{eq:KinVic00}
\nu=\nu_0 (R/R_0)^n,
\ee
where $n\in (0;2)$ is a model parameter, the solution of the differential equation simplifies significantly. The corresponding Green functions for the surface density were found for a few particular cases (see \citealt{1974MNRAS.168..603L,2011MNRAS.410.1007T,2015ApJ...804...87L}). 
The typical time scale of processes in the accretion disc which arise out of viscous diffusion is the viscous time scale
\be
t_{\rm v}(R)\approx \frac{2R^2}{3 \nu(R)}.
\ee
If the kinematic viscosity is described by equation (\ref{eq:KinVic00}) the viscous time scale $t_{\rm v}\propto R^{2-n}$. 
It is useful to introduce the corresponding viscous frequency
\be\label{eq:viscous_frequency}
f_{\rm v}(R)\equiv \frac{1}{t_{\rm v}(R)}. 
\ee 
In case of viscosity prescription given by (\ref{eq:KinVic00}) the viscous frequency $f_{\rm v}\propto R^{n-2}$.

If the viscosity in the accretion disc is described by constant $\alpha$-parameter, it puts constraints on the power-law index $n$ in the description of kinematic viscosity (\ref{eq:KinVic00}). According to equation (\ref{eq:AlphaPrescrNu}) the kinematic viscosity can be represented as $\nu= \alpha\Omega_{\rm K}R^2(H/R)^2$. If the gas pressure and Kramer opacity dominate in the accretion disc, then $(H/R)\propto R^{1/8}$ \citep{2007ARep...51..549S}. As a result, the kinematic viscosity $\nu\propto R^{3/4}$ and therefore $n=3/4$ for the case of constant $\alpha$-parameter (although we note that there is no a priori reason to assume constant $\alpha$). 
The viscous frequency is related to the Keplerian frequency $f_{\rm K}=\Omega_{\rm K}/(2\pi)$ as 
\be\label{eq:frequencies}
f_{\rm v}(R)=3\pi\alpha f_{\rm K}(R)\left(\frac{H(R)}{R}\right)^2. 
\ee

\subsection{Lynden-Bell/Pringle Green functions}
\label{sec:LB}

\begin{figure*}
\centering 
\includegraphics[width=8cm]{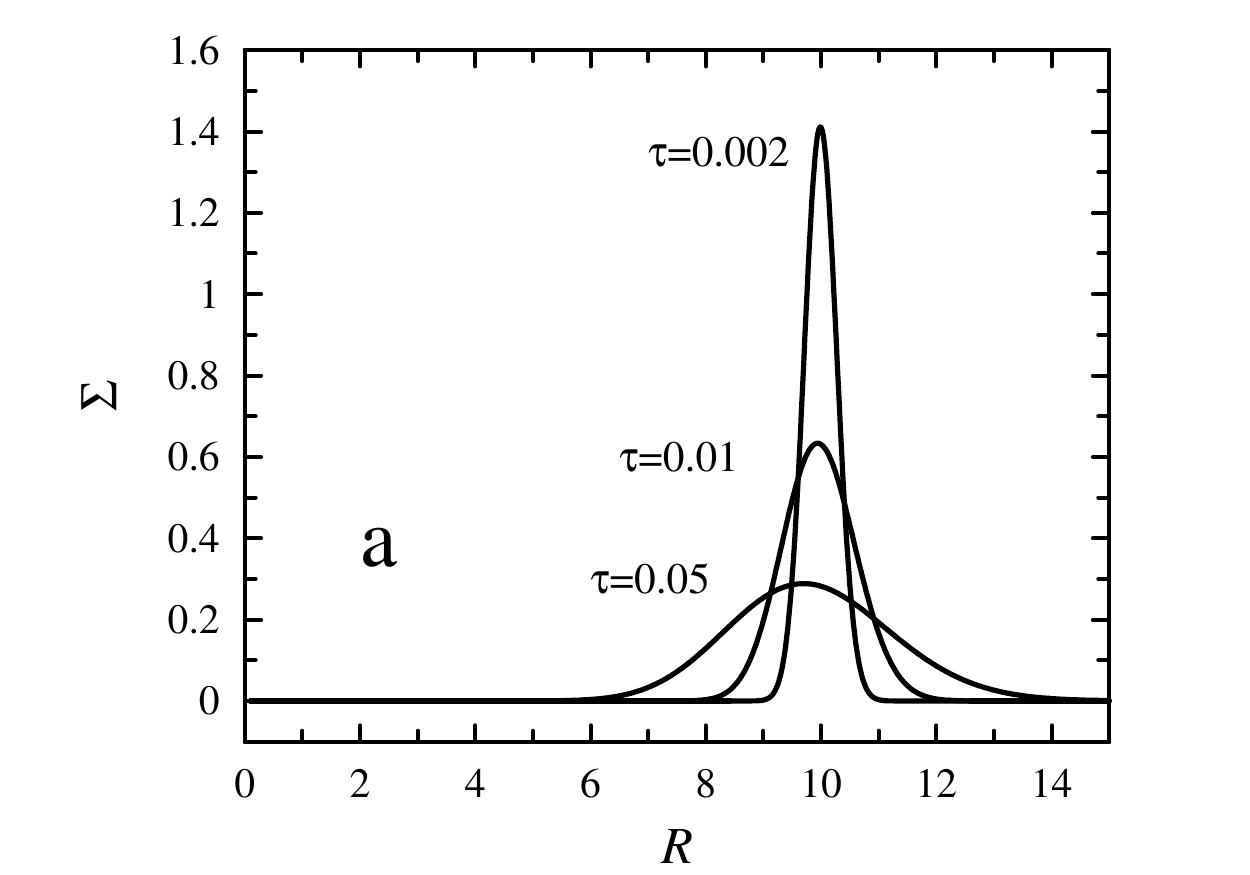} 
\includegraphics[width=8cm]{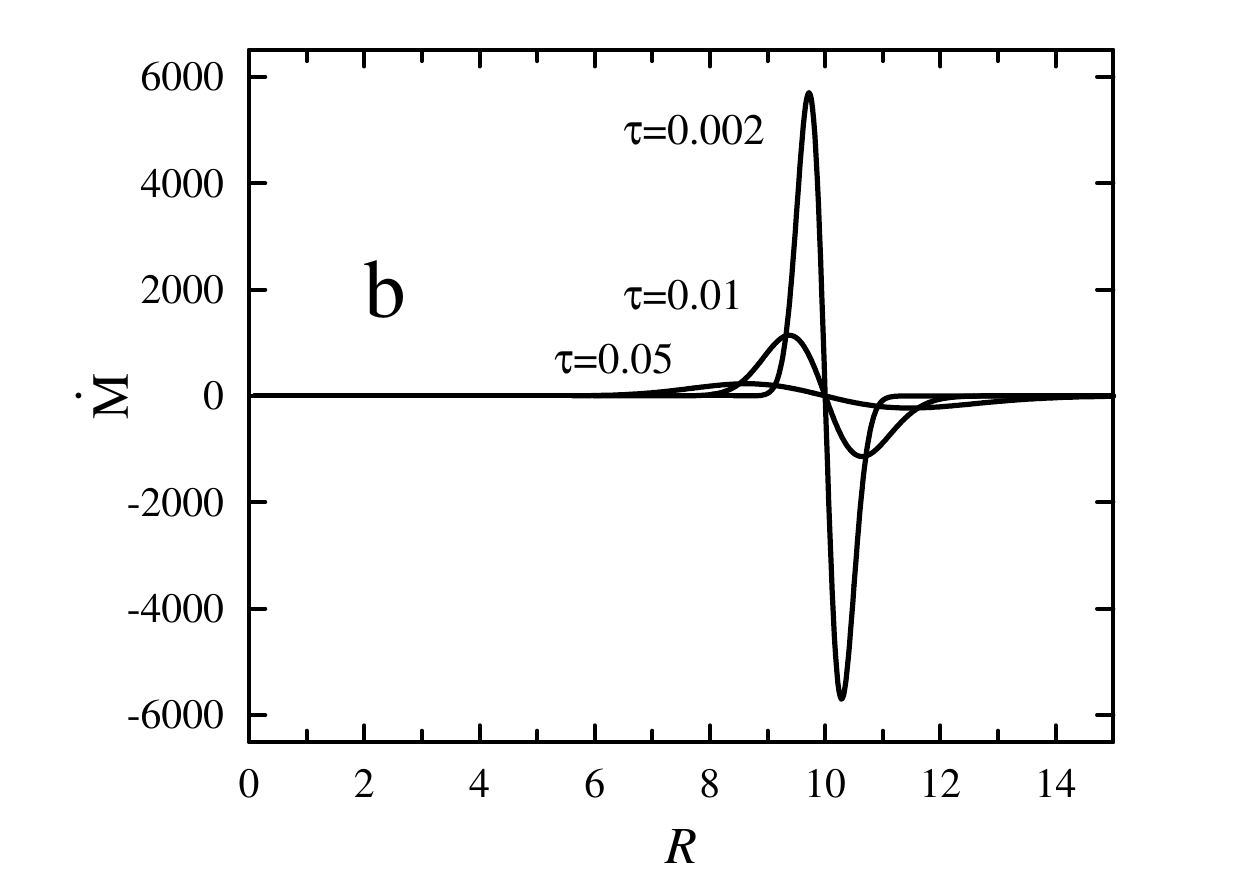} 
\caption{Evolution of (a) surface density $\Sigma$ and (b) local mass accretion rate $\dot{M}$. The inner radius of the disc is taken to be $R_{\rm in}=0$, the surface density at $t=0$ is given by $\delta$-function at $R'=10$. Different curves are given for various time intervals measured in viscous time at radial coordinate $R=1$ ($\tau\equiv t/t_{\nu}(R=1)$): $\tau=0.002,\,0.01$ and $0.05$. It is clearly seen that fluctuations of the surface density and mass accretion rate propagate inwards and outwards. Parameter $n=1$.}
\label{pic:GfTanaka} 
\end{figure*}

\begin{figure*}
\centering 
\includegraphics[width=8cm]{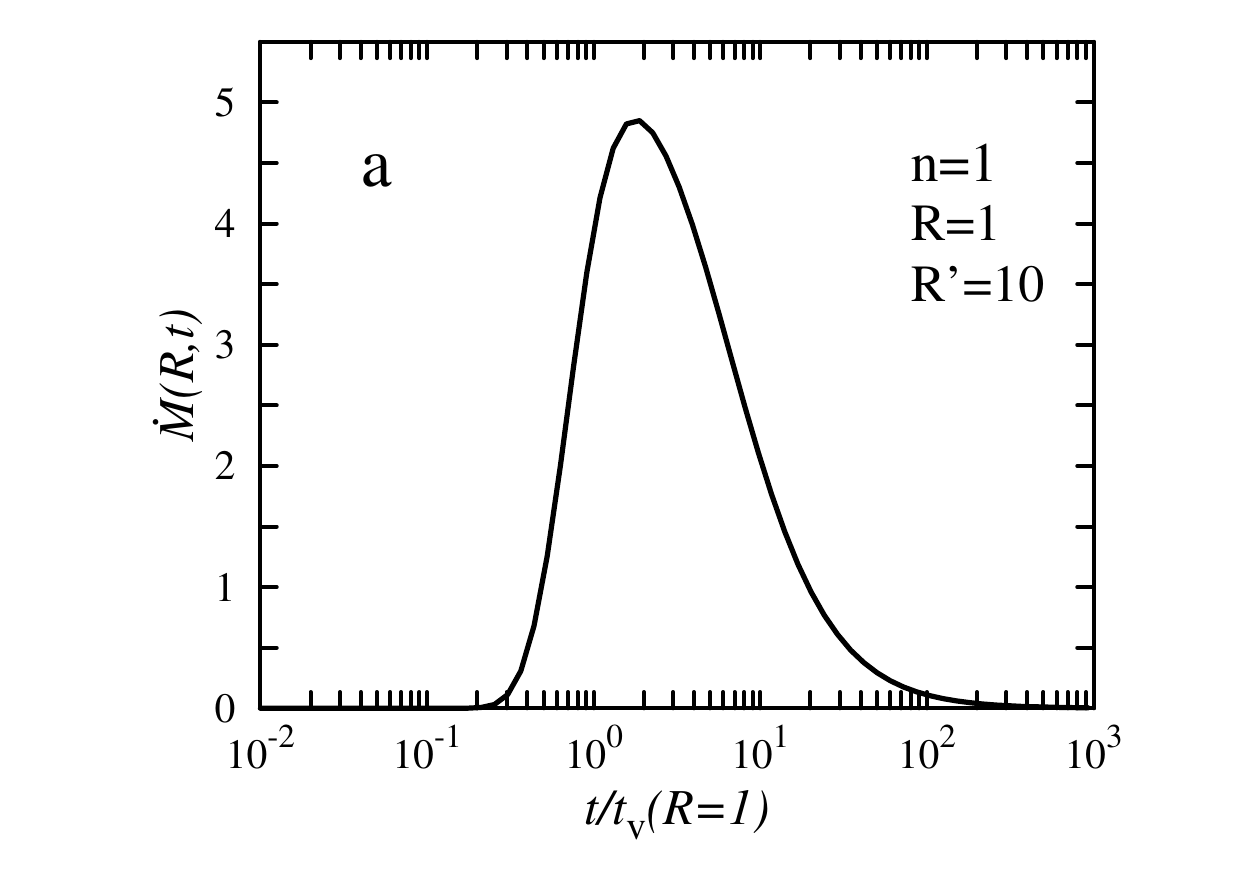} 
\includegraphics[width=8cm]{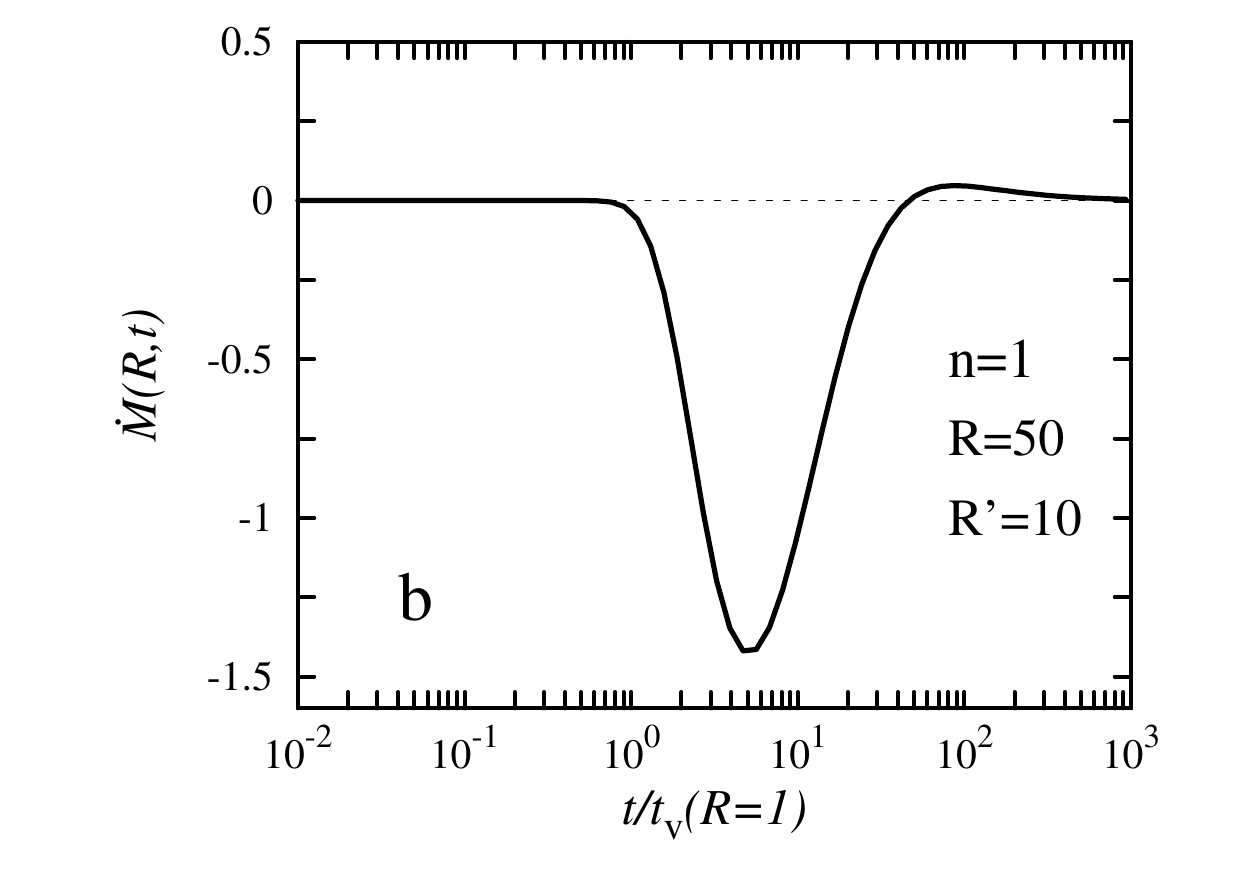} 
\caption{The mass accretion rate caused by viscous diffusion of infinitely thin ring, which surface density at $t=0$ is given by $\delta$-function located at $R'=10$, as a function of dimensionless time (a) at radius $R=1$ and (b) at radius $R=50$. The inner disc radius is taken at $R_{\rm in}=0$. At radius $R=50$, which outside  radius $R'=10$, the mass accretion rate is negative initially, but it becomes positive after sufficiently large time. Parameter $n=1$.}
\label{pic:Mdot(t)} 
\end{figure*}

For the case of $R_{\rm in}=0$ and $R_{\rm out}=\infty$ the Green function for the surface density was found by \cite{1974MNRAS.168..603L}. The Green function is given by
\beq
G(R,R',\tau)=\frac{1}{2(2-n)}\tau^{-1}R_c^{-1}\left(\frac{R'}{R_c}\right)^{1-n}\left(\frac{R'}{R}\right)^{n+1/4} \nonumber \\
\times\exp\left[-\frac{2l^2}{\tau}\left( \left(\frac{R'}{R_{\rm c}}\right)^{2-n}+ \left(\frac{R}{R_{\rm c}}\right)^{2-n}  \right)\right] \nonumber \\
\times I_{l}\left(\frac{4l^2}{\tau}\left(\frac{RR'}{R^2_{\rm c}}\right)^{\frac{2-n}{2}}\right),
\eeq
where $\tau=t/t_{\rm v}$ is time measured in units of viscous time at fiducial radius $R_{\rm c}$, $I_{\nu}$ is a modified Bessel function of the first kind and $l=(4-2n)^{-1}$. 

According to equation (\ref{eq:mdot02}), the Green function for the mass accretion rate is
\beq
\label{eq:Gf_MdotLB}
G_{\dot{M}}(R,R',\tau)=
6\pi\nu_0 l^2 \tau^{-2} R^{\frac{3-2n}{4}} R ^{\prime \frac{9-2n}{4}} R_{\rm c}^{2n-2} \nonumber \\
\times \exp\left[-\frac{2l^2}{\tau} \left( \left(\frac{R'}{R_{\rm c}}\right)^{2-n}+ \left(\frac{R}{R_{\rm c}}\right)^{2-n}  \right) \right] \nonumber \\
\times \left\{
I_{l-1}\left(\psi \right)-I_{l}\left( \psi \right)\left(\frac{R}{R'}\right)^{1-n/2}
\right\},
\eeq
where
$$\psi=\frac{4l^2}{\tau}\left(\frac{RR'}{R^2_{\rm c}}\right)^{\frac{2-n}{2}}.$$
It is interesting to note a symmetry property of the mass accretion rate Green function:
\be\label{eq:dotMsym}
G_{\dot{M}}(R,R',\tau)=a^{1-n}G_{\dot{M}}(aR,aR',a^{2-n}\tau),
\ee
where $a$ is an arbitrary constant. 

If the argument of the Bessel functions in (\ref{eq:Gf_MdotLB}) is small enough, which is the case at sufficiently large $\tau$:
$$\tau\gg 4l^2\left(\frac{\max(R,R')}{R_{\rm c}}\right)^{2-n}= 4l^2\frac{\max(t_{\rm v}(R),t_{\rm v}(R'))}{t_{\rm v}(R_{\rm c})},$$
one can use the approximation:
\beq
G_{\dot{M}}(R,R',\tau)\approx
6\pi\nu_0 l^2 R^{\frac{3-2n}{4}} R^{\prime\frac{9-2n}{4}} R_{\rm c}^{2n-2} \nonumber \\
\times\frac{\tau^{-(l+1)}}{\Gamma(l)}
\left[2l^2 \left(\frac{R R'}{R^2_{\rm c}}\right)^{1-n/2}\right]^{l-1},
\eeq
where $\Gamma(x)$ is the gamma function.

According to (\ref{eq:Gf_MdotLB}) the mass accretion rate caused by surface density concentrated at $R'$ at $t=0$ is positive in the inner disc region ($R<R'$). In the outer regions ($R>R'$)  the Green function (\ref{eq:Gf_MdotLB}) gives a negative mass accretion rate at first, 
\footnote{Negative mass accretion rate {(i.e. mass transfer from smaller to larger radii)} is to be understood as the disturbance on top of a, typically larger, positive average mass accretion rate $\dot{M}_0$, i.e., one resulting in a smaller, but still positive, mass accretion rate. }
which takes away the angular momentum of the inner parts of the disc. However, after a sufficiently long time the mass accretion rate at a given radius $R>R'$ becomes positive. The change of sign of the mass accretion rate happens at approximately the viscous time scale corresponding to the radius $R$ (see Appendix \ref{App:Mdot}).

\subsection{Green function for truncated disc with given inner radius}

In a more general case with inner radius $R_{\rm in}$ \citep{2011MNRAS.410.1007T}, zero torque at $R_{\rm in}$ and $R_{\rm out}=\infty$ the Green functions are given by
\beq\label{eq:GfTanaka}
& G(R,R',t)=\left(1-\frac{n}{2}\right)R^{-n-1/4}R'^{5/4}R_{\rm in}^{n-2} & \\
&
\times\int\limits_{0}^{\infty}[J_l(kx)Y_l(k)-Y_l(kx)J_l(k)][J_l(kx')Y_l(k)-Y_l(kx')J_l(k)]& \nonumber \\
&
\times(J_l^2(k)+Y_l^2(k))^{-1}
\nonumber \\
&
\times\exp\left[-2\left(1-\frac{n}{2}\right)^2 k^2\frac{t}{t_{{\rm v,\,in}} }\right]k\d k,
& \nonumber
\eeq
where $x=(R/R_{\rm in})^{1-n/2}$, $t_{\rm v,\, in}=\frac{2}{3}R_{\rm in}^2/\nu(R_{\rm in})$ is local viscous time at the inner disc radius and $J_{l}(x)$ and $Y_l(x)$ are the Bessel functions of the first and second kind. The corresponding Green function for the mass accretion rate can be found according to equation (\ref{eq:mdot02}).

The Green functions for the case of $R_{\rm in}=0$ and $R_{\rm out}<\infty$ were found by \cite{2015ApJ...804...87L}.

The exact boundary conditions affect the typical time scales in the accretion disc (see Appendix\,\ref{App:Mdot}). The influence of advanced Green functions (with boundary conditions given at arbitrary radial coordinates and derived under the assumption of non-Newtonian gravitational potential) will be explored in further works.

\section{Exploring properties of the Green function}

In this section, we explore the properties of the Lynden-Bell \& Pringle Green function introduced in the previous Section. The properties of a more general Green function (for the case of given inner or outer radius) are qualitatively similar.

\subsection{Time domain}

\begin{figure}
\centering 
\includegraphics[width=8cm]{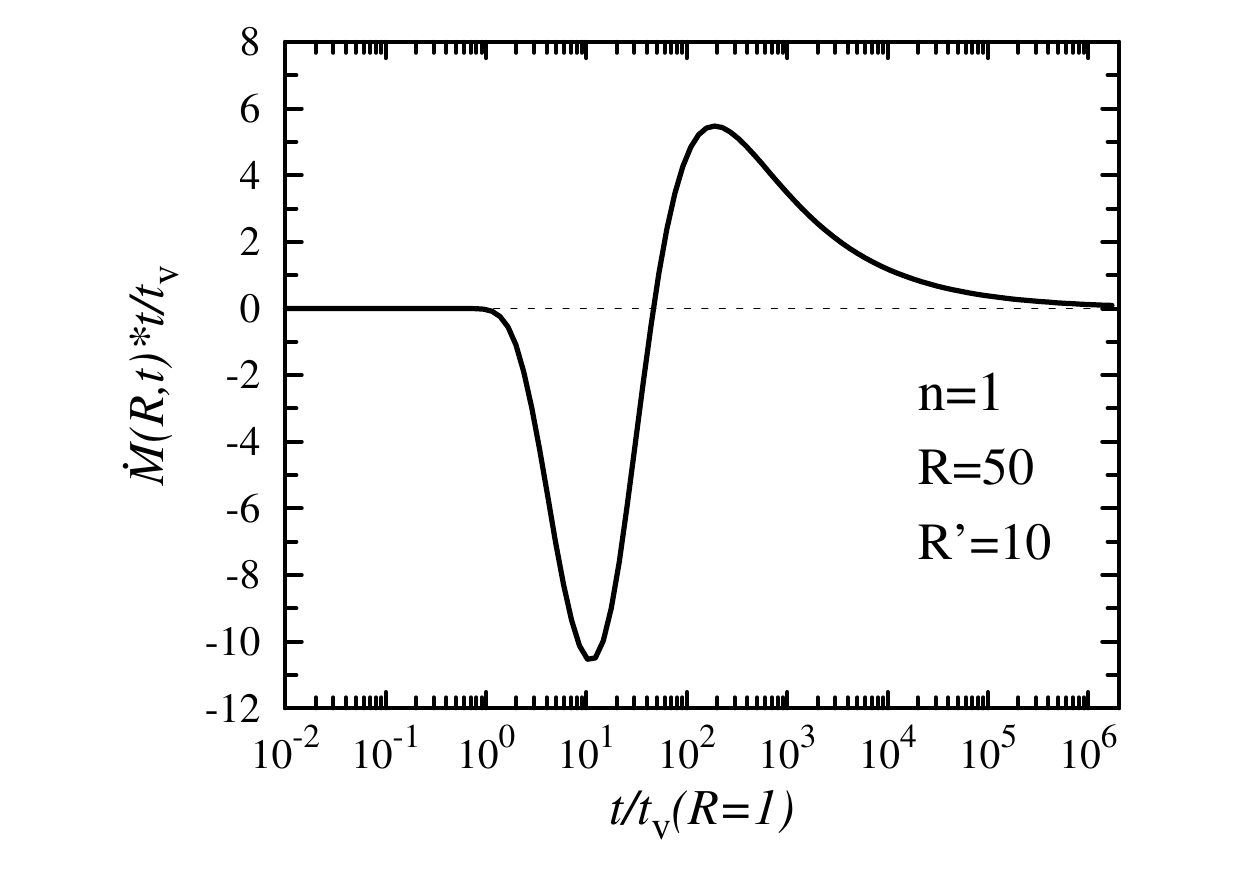} 
\caption{The mass accretion rate caused by surface density perturbation at $R'=10$ at $t=0$ as a function of dimensionless time at radius $R=50$. The inner disc radius is taken at $R_{\rm in}=0$. One can see that almost all material accretes inside after sufficient amount of time. Parameter $n=1$.}
\label{pic:Mdot(t)_} 
\end{figure}

The Green function of the surface density in the time domain $G(R,R',t)$ describes the evolution of the surface density given initially by a $\delta$-function at radial coordinate $R'$: $\Sigma(R,t=0)=\delta(R-R')$. The viscous force produces an angular momentum exchange between adjacent rings, such that the fast inner rings pass their angular momentum to slower outer rings. As a result, a part of the material accretes inwards losing angular momentum while another part moves outwards carrying out angular momentum. This process leads to spreading of an initially narrow ring (see Fig.\,\ref{pic:GfTanaka}a). The corresponding evolution of the mass accretion rate is given by the Green function for the mass accretion rate $G_{\dot{M}}(R,R',t)$ (see Fig.\,\ref{pic:GfTanaka}b). Spreading of the initial ring is manifested initially by a positive mass accretion rate at $R<R'$ (see Fig.\,\ref{pic:Mdot(t)}a) and a negative mass accretion rate at $R>R'$ (see Fig.\,\ref{pic:Mdot(t)}b). The radial coordinate where the mass accretion rate changes its sign moves outwards with time, which means that the angular momentum is picked up finally by a small fraction of the material, while most of it accretes onto the central object (see Fig.\,\ref{pic:Mdot(t)_}). 

We apply Green functions in order to describe the evolution of fluctiations arising on top of the average accretion flow. {Therefore, the negative mass accretion rates predicted by the Green functions results only in lower mass accretion rate at certain regions of the accretion disc, rather than genuinely negative accretion rates.
\footnote{Note that we consider the case of linear equation of viscous diffusion.}
The diffusion of a perturbation requires transfer of angular momentum from the inner parts of the disc to its outer parts. The extra angular momentum supplied from the inner parts of the disc slightly slows down the local accretion process and results in lower mass accretion rate.}

\subsection{Frequency domain}

The evolution of mass accretion rate in the time domain is described by convolution of the Green function and the initial fluctuations of the surface density (\ref{eq:mdot01_}). 
The convolution turns to production in the frequency domain, which simplifies the problem. It makes the properties of Green functions in the frequency domain essentially important.

The Green function in the frequency domain is obtained by the Fourtier transform
\be
\overline{G}_{\dot{M}}(R,R',f)=\int\limits_{-\infty}^{\infty}\d x\,G_{\dot{M}}(R,R',x)\,e^{-2\pi i f x}. 
\ee
{The obtained Green function shapes variability that originated from radius $R'$, at given radius $R$.}

The qualitative behavior of the Green function in the frequency domain is different for perturbations propagating initially inwards ($R<R'$) and outwards ($R>R'$).

\begin{figure*}
\centering 
\includegraphics[width=16cm]{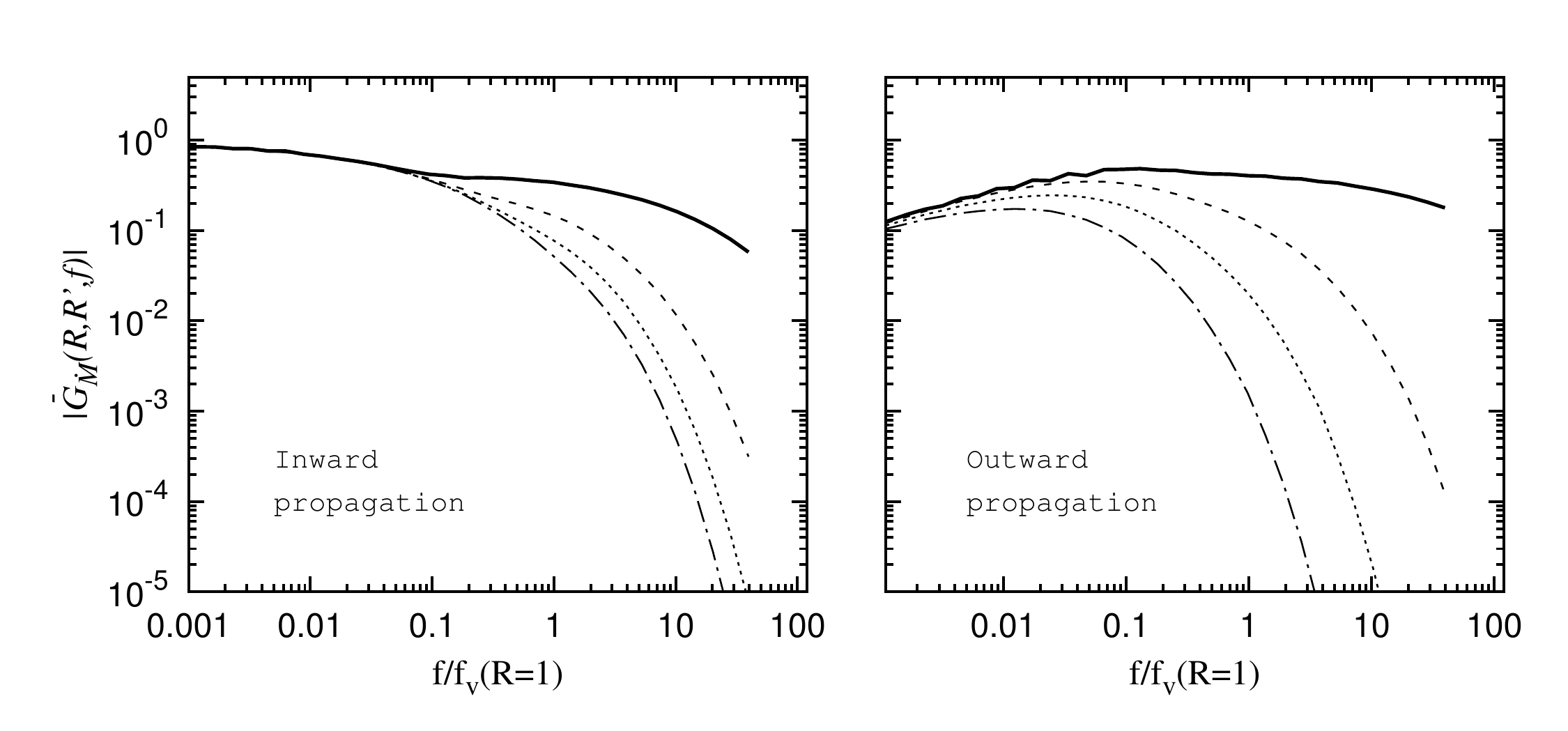} 
\caption{The absolute value of Green function of the mass accretion rate in the frequency domain. The inner disc radius is located at $R_{\rm in}=0$, the initial perturbation of the surface density is at $R'=10$. The radial coordinates of final perturbations are taken to be $R=1,\,2,\,4,\,8$ (solid, dashed, dotted and dashed-dotted lines respectively), i.e. perturbations propagate inwards (left) and $R=11,\,20,\,40,\,80$ (solid, dashed, dotted and dashed-dotted lines respectively), i.e. perturbations propagate outwards (right). The parameter $n=1$.}
\label{pic:PS13} 
\end{figure*}

\begin{figure*}
\centering 
\includegraphics[width=16cm]{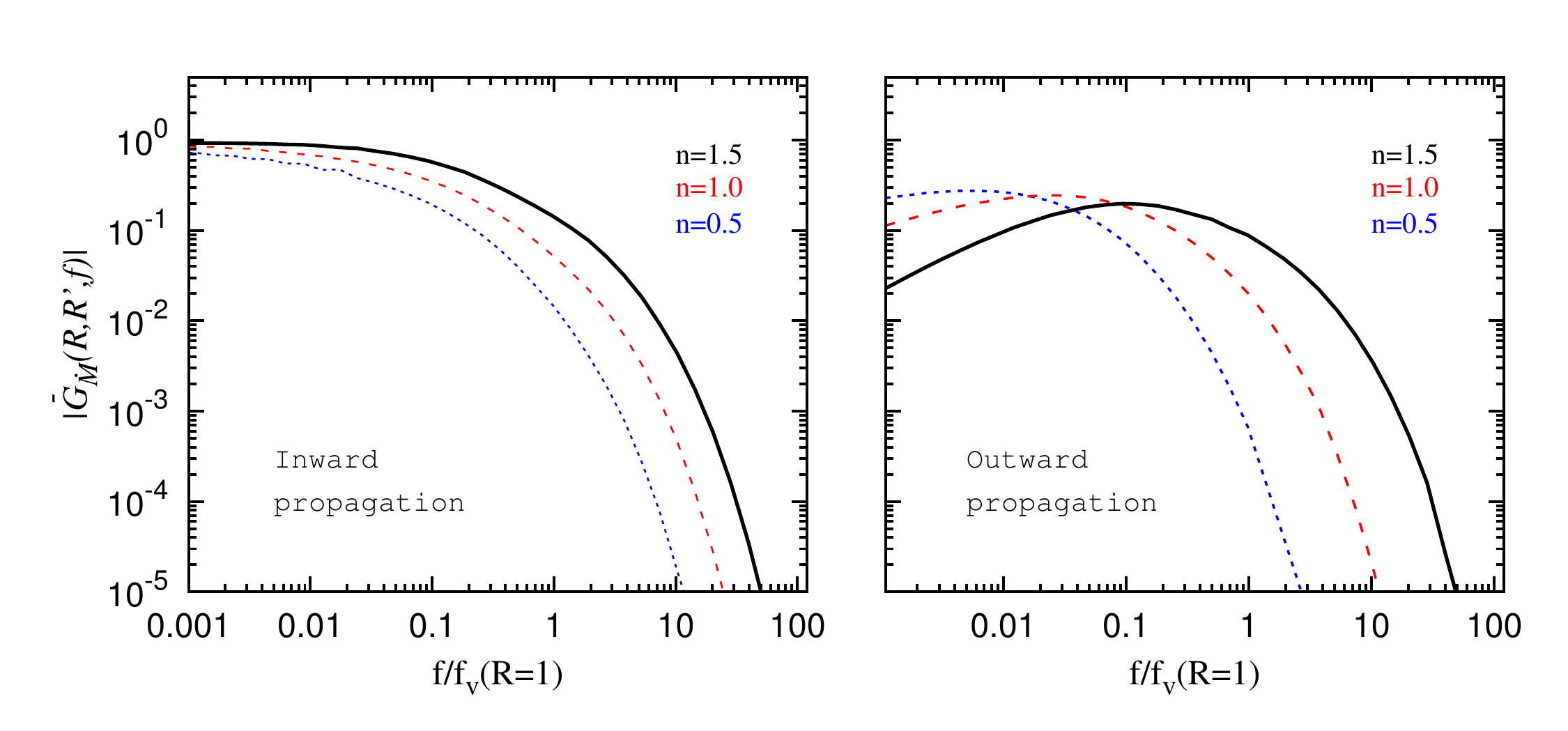} 
\caption{The absolute value of the Green function of the mass accretion rate in the frequency domain. The inner disc radius is fixed at $R_{\rm in}=0$, the initial perturbations are at $R'=10$. The radial coordinate of final variability is at $R=1$ (left) and $R=40$ (right). Black solid, red dashed and blue dotted lines are given for $n=1.5,\,1$ and $0.5$ respectively. Suppression of variability at high frequencies is stronger for lower parameter $n$. Suppression of variability at low frequencies for outwards propagation is stronger for higher parameter $n$.}
\label{pic:PS18} 
\end{figure*}

In the case of inwards propagation, the absolute value of the Green function is $|\overline{G}_{\dot{M}}|\approx 1$ for frequencies much lower that the local viscous frequency $f_{\rm v}(R)$, and is suppressed for frequencies above $f_{\rm v}(R)$ (see Fig.\,\ref{pic:PS13} \textit{left}). It results in suppression of high frequency variability originating from distant outer radii. The suppression at high frequencies depends on the separation between radial coordinates of initial and final fluctuations ($R'$ and $R$): the larger the separation, the stronger Green function suppression. The suppression is different for different viscosity prescriptions given by the parameter $n$: the lower $n$ is, the larger the range of viscous time scales over the radial coordinates in the disc is, the stronger the suppression of the Green function (see Fig.\,\ref{pic:PS18} \textit{left}).

In the case of fluctuations propagating outwards ($R>R'$) the absolute value of the Green function $|\overline{G}_{\dot{M}}|$ is suppressed at high frequencies and at low frequencies, reaching its maximum at frequency close to local viscous frequency $f_{\rm v}(R')$ (see Fig.\,\ref{pic:PS13} \textit{right}). This property is caused by the fact that initially negative mass accretion rate at $R>R'$ turns to zero after a while and becomes positive on a sufficiently long timescale. The typical timescale of this process is close to the local viscous time scale (see Appendix\,\ref{App:Mdot}), which results in maximum of the $|\overline{G}_{\dot{M}}|$ being at $f_{\rm v}(R')$ (see Fig,\,\ref{pic:PS18} \textit{right}).

\begin{figure*}
\centering 
\includegraphics[width=18cm]{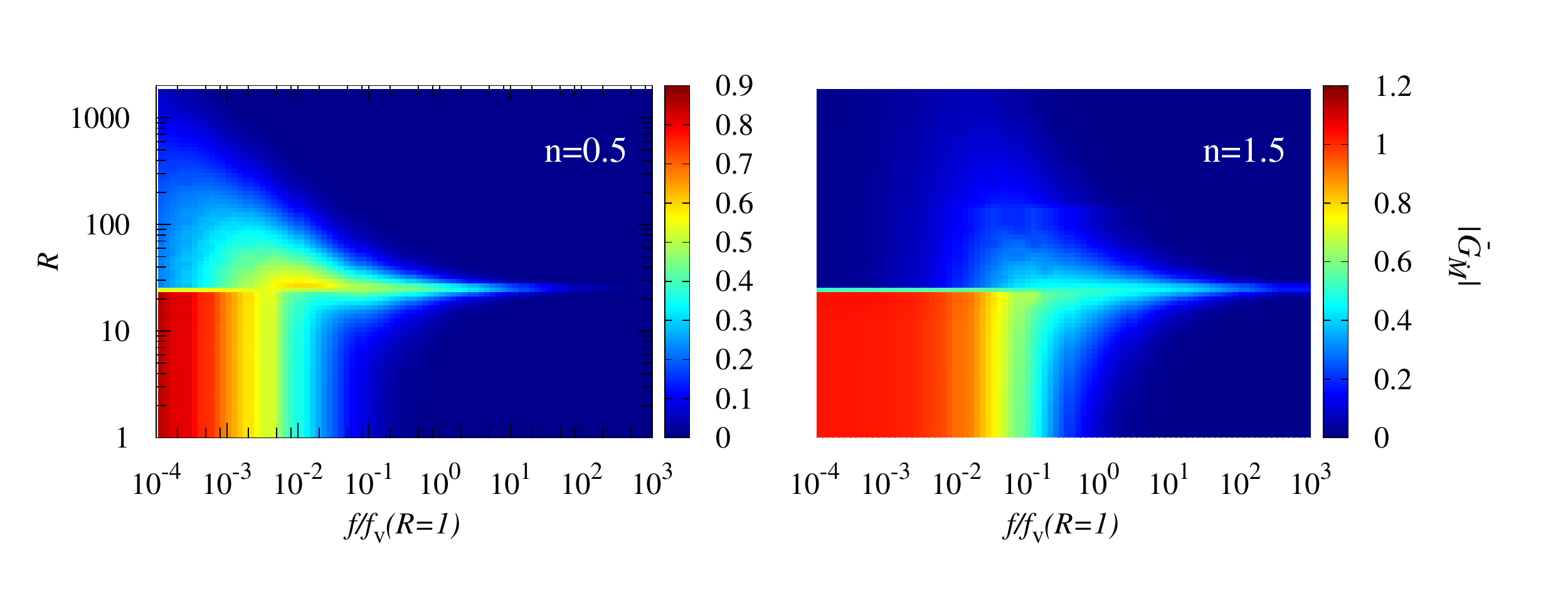}
\caption{The absolute value of the Green  function for mass accretion rate in the frequency domain as a function of final radial coordinate $R$ and dimensionless frequency  $f/f_{\rm v}(R=1)$. The results are given for parameters $n=0.5$ (left) and $n=1.5$ (right). The initial perturbation is given by $\delta$-function and located at radial coordinate $R'=25$.}
\label{pic:14_LB} 
\end{figure*}

The absolute value of the Green function for outwards propagation at its maximum can be comparable to that for inward propagation (see Fig.\,\ref{pic:14_LB}). In this case the influence of the inner parts of the accretion disc on the local mass accretion rate variability can dominate over the influence of the outer parts of the disc. However, the influence of inner disc applies only to the closest radial coordinates and strongly is suppressed for distant outer regions (see Fig.\,\ref{pic:PS13} \textit{right}).

\begin{figure*}
\centering 
\includegraphics[width=18cm]{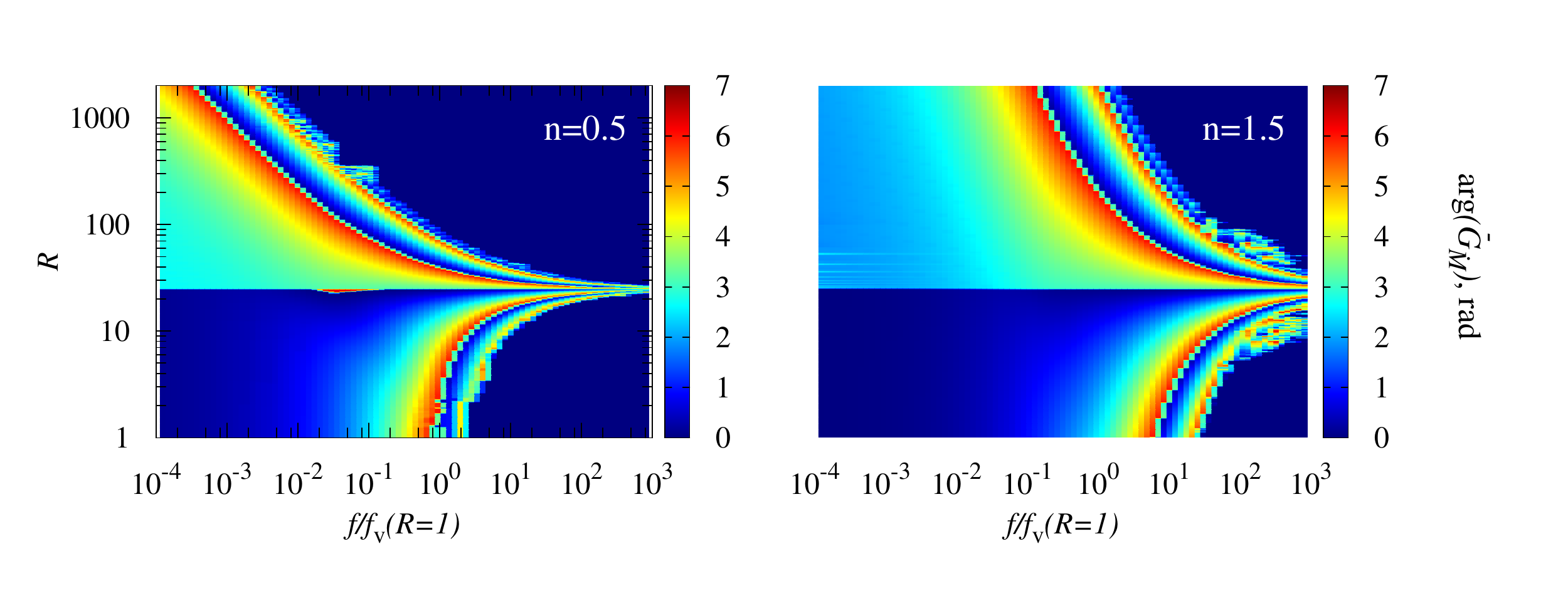}
\caption{The phase angle of the Green function for the mass accretion rate in the frequency domain as a function of radial coordinate $R$ and dimensionless frequency $f/f_{\rm v}(R=1)$. The results are given for for parameters $n=0.5$ (left) and $n=1.5$ (right). The initial perturbation is given by $\delta$-function and located at $R'=25$.}
\label{pic:GF_phase} 
\end{figure*}

In Fig.\,\ref{pic:GF_phase}, we plot the phase angle in radians of the Green function for two values of $n$. For low frequencies ($f<<f_v$), we see that $\dot{M}(R<R',t)$ lags behind $\dot{M}(R=R',t)$, which simply represents the inwards propagation of the density perturbation at $R'$. Conversely, $\dot{M}(R>R',t)$ \textit{leads} $\dot{M}(R=R',t)$ (note that a phase angle $>\pi$ and $<2\pi$ is indistinguishable from a phase angle $<0$ and $>-\pi$), because the accretion rate upstream of the perturbation is negative (e.g. Fig.\,\ref{pic:Mdot(t)}b). For higher frequencies, we see that phase wrapping causes oscillations in the phase angle.

According to (\ref{eq:dotMsym}) the Lynden-Bell Green function (see Section\,\ref{sec:LB}) has the property:
\beq
\overline{G}_{\dot{M}}(R,R',f)=a^{-1}\overline{G}_{\dot{M}}\left(aR,aR',\frac{f}{a^{2-n}}\right),
\eeq
where $a>0$ is an arbitrary constant. Therefore the Green function in this particular case can be considered as a function of two variables instead of three (i.e. no need to consider a different $R'$ value to what we already plotted in Fig.\,\ref{pic:14_LB}).

\section{The propagating fluctuations model using Green functions}

In the propagating fluctuation model, the initial variability of mass accretion rate is assumed to be stirred up through out the disc, i.e. fluctuations are introduced at a \textit{range} of $R'$. These fluctuations propagate to other parts of the accretion disc and, in general, interact (multiplicatively) with one another. In previous treatments of the model, only inwards propagation has been considered and/or a very simple Green function has been assumed to describe propagation \citep{1997MNRAS.292..679L,2001MNRAS.327..799K,2013MNRAS.434.1476I}.
Here, we consider both inwards and outwards propagation of fluctuations. In the simplest case, we may consider the fluctuations to simply pass one another. In this case, linearity is preserved. In general though, fluctuations interact with one another as they propagate: e.g. inward propagating fluctuations from a large $R'$ will modulate and be modulated by outward propagating fluctuations from a smaller $R'$. This general case is non-linear.

\subsection{Mass accretion rate fluctuations: the general case}

The total mass accretion rate consists of the average mass accretion rate $\dot{M}_0$ and fluctuations $\dot{m}(R,t)$ on top of the average value:
\be
\dot{M}(R,t)=\dot{M}_0+\dot{m}(R,t). 
\ee
We assume that local perturbations of surface density $A(R,t)$ caused by MHD are $A(R,t)\propto \dot{M}(R,t)/R$. 

Indeed, the evolution of surface density is given by
\footnote{{Using the following designation for the partial derivative: $\partial_x\equiv \partial/\partial x$.}}
$$R\partial_t\Sigma=\partial_R\dot{M}.$$ 
The local mass accretion rate can be represented as $\dot{M}=2\pi R v_R\Sigma$, where $v_R$ is local radial velocity. The radial velocity is defined by the diffusion process and can be disturbed by local interaction of the accretion flow with magnetic field. Then we can represent the radial velocity as
$$
v_r(R,t)=v_r^{\rm (d)}(R,t)[1+\vartheta(R,t)],
$$
where $v_r^{\rm (d)}(R,t)$ is the radial velocity defined by the diffusion process and $\vartheta(R,t)\ll 1$ represents fluctuations of radial velocity caused by local MHD processes. Then the local surface density evolution is defined by the relation {
\beq
R\partial_t \Sigma&=&\partial_R\left\{R\Sigma v_r^{\rm (d)}(R,t)[1+\vartheta(R,t)]\right\}\nonumber\\
&=&\partial_R\left\{\dot{M}^{\rm (d)}(R,t)[1+\vartheta(R,t)]\right\},\nonumber
\eeq
where $\dot{M}^{\rm (d)}(R,t)$ is the mass accretion rate caused by viscous diffusion only. Differentiation over the radial coordinate gives
\beq
R\partial_t \Sigma &=&\partial_{R}\dot{M}^{\rm (d)}(R,t)\left[1+\vartheta(R,t)\right]+\dot{M}^{\rm (d)}(R,t)\partial_R\vartheta(R,t) \nonumber \\
&\approx&\partial_{R}\dot{M}^{\rm (d)}(R,t)+\dot{M}^{\rm (d)}(R,t)\partial_R\vartheta(R,t). 
\eeq
The first term in the right-hand side (rhs) of relation defines variations of the surface density due to the diffusion process, while the second term defines new perturbations due to local fluctuations of viscosity. Hence, the initial perturbations of the radial density $R\Sigma(R)$ are proportional to local mass accretion rate: $\partial_t\Sigma^*\propto \dot{M}/R$. Therefore, the perturbations of the surface density can be represented as follows:}
\be\label{eq:LocPert01}
A(R,t)=\frac{a(R,t)}{R}\frac{\dot{M}(R,t)}{\dot{M}_0}=\frac{a(R,t)}{R}\left(1+\frac{\dot{m}(R,t)}{\dot{M}_0}\right),  
\ee
where $a(R,t)$ is defined by local physics of interaction between accretion flow and magnetic field and does not depend on conditions in other parts of accretion disc.
Then we get a non-linear equation for the perturbations of the mass accretion rate in the time domain:
\be
\label{eq:mdot_tdomain}
\dot{m}(R,t)=\int\limits_{R_{\rm in}}^{R_{\rm out}}\d R'\,G_{\dot{M}}(R,R',t)\otimes_t
 \frac{a(R',t)}{R'}\left(1+\frac{\dot{m}(R',t)}{\dot{M}_0}\right)
\ee

The local perturbations of the surface density in the frequency domain are given by 
\be\label{eq:LocalPertA}
\overline{A}(R,f)=
\frac{\overline{a}(R,f)}{R}\otimes_f
\left(\delta(f)+\overline{\frac{\dot{m}(R,f)}{\dot{M}_0}}\right),
\ee 
where $\otimes_f$ denotes a convolution in frequency.

The local perturbations of surface density propagate through the accretion disc according to the solution of the diffusion equation (\ref{eq:DifEqGen}) and define the final perturbations of the mass accretion rate. If the final mass accretion rates are given by convolution according to equation (\ref{eq:mdot01}) the perturbations of mass accretion rate in the frequency domain are given by
\beq\label{eq:NoiceMult0}
\overline{\dot{m}}(R,f)=
\int\limits_{R_{\rm in}}^{R_{\rm out}}\d R'\,\overline{G}_{\dot{M}}(R,R',f)\overline{A}(R',f),
\eeq
where $\overline{G}_{\dot{M}}(R,R',f)$ is the corresponding Green function in the frequency domain and $R_{\rm in}$ and $R_{\rm out}$ are inner and outer radii of accretion disc. 
\footnote{The inner and outer radii in equation (\ref{eq:NoiceMult}) are defined by the accretion disc geometry and the physical conditions in the disc. The outer radius $R_{\rm out}$ cannot exceed the radius where the effective temperature of accretion disc is lower than $\simeq 6500\,{\rm K}$. Otherwise the accretion disc becomes thermally unstable because of hydrogen recombination and a dramatic change of opacity \citep{2001NewAR..45..449L}. Moreover, the initial perturbations of the mass accretion rate due to MHD processes are expected to be weaker in case of low temperature and weakly ionized plasma.}
Substitution of expression (\ref{eq:LocalPertA}) into (\ref{eq:NoiceMult0}) gives
\beq\label{eq:NoiceMult}
&\overline{\dot{m}}(R,f)&=\int\limits_{R_{\rm in}}^{R_{\rm out}}\frac{\d R'}{R'}\,\overline{G}_{\dot{M}}(R,R',f)\overline{a}(R',f)+ \\
&&\frac{1}{\dot{M}_0}\int\limits_{R_{\rm in}}^{R_{\rm out}}\frac{\d R'}{R'}\,
\overline{G}_{\dot{M}}(R,R',f)
\left[\overline{a}(R',f)\otimes_f\overline{\dot{m}}(R',f)\right].  \nonumber
\eeq
If the perturbations of the mass accretion rate are significantly smaller than the average mass accretion rate ($\dot{m}(R,t)\ll\dot{M}_0$) then the perturbations of mass accretion rate are defined by the first term on the rhs of equation (\ref{eq:NoiceMult}). In this case they are represented by a superposition of fluctuations originating at different radii and modified by the diffusion process. 
If perturbations of the mass accretion rate are comparable with the average mass accretion rate then the final result is affected by both terms of the rhs of equation (\ref{eq:NoiceMult}).

\subsection{The equation in discrete form: particular case of Green function given by $\delta$-function}

In order to directly compare our formalism with that of \cite{2013MNRAS.434.1476I}
let us consider how the formalism transforms if the flow is split into $N$ rings of radii $R_N<R_{N-1}<...<R_2<R_1$. Then the equation (\ref{eq:mdot_tdomain}) takes form
\be
\dot{m}(R_i,t)\simeq\sum \limits_{j>i} \Delta R_j\,G(R_i,R_j,t)\otimes_t
 \frac{a(R_j,t)}{R_j}\left(1+\frac{\dot{m}(R_j,t)}{\dot{M}_0}\right).\nonumber
\ee
If the Green function is given by $\delta$-function (the case of arbitrary Green function is discussed in Appendix \ref{Sec:EqDiscForm2}): $G(R_i,R_j,t)=\delta(t-\Delta t_{ij})$, where $\Delta t_{ij}$ is the propagation time from $R_j$ to $R_i$, we get
\be
\dot{m}(R_i,t)\simeq\sum \limits_{j>i} \Delta R_j\,\frac{a(R_j,t-\Delta t_{ij})}{R_j}\left(1+\frac{\dot{m}(R_j,t-\Delta t_{ij})}{\dot{M}_0}\right).\nonumber
\ee
Then the total mass accretion rate at the ring $i$ can be represented as follows:
\beq 
& \dot{M}(R_i,t)=\dot{M}(R_{i-1},t-\Delta t_{i,(i-1)})&   \nonumber  \\
&+ \Delta R_{i-1}\,\frac{a(R_{i-1},t-\Delta t_{i,(i-1)})}{R_{i-1}}\left(1+\frac{\dot{m}(R_{i-1},t-\Delta t_{i,(i-1)})}{\dot{M}_0}\right)&
\nonumber  
\eeq 
or
\beq
&\dot{M}(R_i,t)=\dot{M}(R_{i-1},t-\Delta t_{i,(i-1)})& \nonumber\\
 &\times\mathcal{A}_{n-1}(t-\Delta t_{i,(i-1)}),&
\eeq
where
\beq
\mathcal{A}_{n-1}(t-\Delta t_{i,(i-1)})=
1+ \frac{\Delta R_{i-1}}{R_{i-1}}\,\frac{a(R_{i-1},t-\Delta t_{i,(i-1)})}{\dot{M}_0}. \nonumber
\eeq
Therefore considering this particular case, the total mass accretion rate in the $i$-th ring can be obtained from the mass accretion rate in the $(i+1)$th ring by multiplying on the stochastic function with a mean of unity. 
Then the total mass accretion rate in the frequency domain can be calculated:
\beq
\overline{\dot{M}}(R_n,f)=\overline{\dot{M}}(R_1)\coprod\limits_{i=1}^{n-1}\overline{\mathcal{A}}_{i}(f)e^{2\pi i f\Delta t_{i,(i-1)}},
\eeq
where $\overline{\mathcal{A}}_{i}(f)$ is the Fourier transform of the function $\mathcal{A}_{i}(t)$. If fluctuations at different radii are not correlated with each other the power of the mass accretion rate is
\beq
|\overline{\dot{M}}(R_n,f)|^2=\overline{\dot{M}}^2(R_1)\coprod\limits_{i=1}^{n-1}|\overline{\mathcal{A}}_{i}(f)|^2,
\eeq
This result agrees with the formalism, which was developed earlier by \cite{2013MNRAS.434.1476I}.

\section{Power spectra of initial perturbations}

We assume that the initial fluctuations of the surface density are stochastic and have a random phase but well-defined power spectrum. This power spectrum depends on the physical process generating variability. 

The initial perturbations might be caused by MHD turbulence, which serves as the source of viscosity in the accretion flow, in a form of fluctuations in magnetic stress \citep{1991ApJ...376..214B,1995ApJ...440..742H,1995ApJ...446..741B}.

The typical time-scale can be defined by the dynamo process \citep{2004MNRAS.348..111K}, which is generally shorter than the viscous time scale. However, the time scale for producing magnetic field sufficiently coherent to affect the accretion rate can be longer and comparable with viscous time-scale \citep{2003ApJ...593..184L}.

An accurate description of the emergence of new fluctuations of the mass accretion rate requires the introduction of their cross-spectrum (see Appendix \ref{App:PSD}), while the PDS of the initial fluctuations at each radial coordinate can be used for approximate description of real process.

Our formalism allows us to use any power spectra for the new fluctuations. Here we will focus only on some possibilities.

\subsection{White noise}

The power spectrum of white noise is given by a constant. Though this kind of process is not physical, it can be used as an approximation in cases when the break frequency of the initial perturbations is well above local viscous frequency.

\subsection{Lorentzian profile}

The power spectrum at each radius can be represented by a zero-centred Lorentzian breaking at frequency $f_{\rm br}(R)$:
\be\label{eq:LorentzProf}
S_a(f)=\left|a(f)\right|^2 = \frac{2 P(R)}{\pi}\frac{f_{\rm br}(R)}{(f_{\rm br}(R))^2+f^2}, 
\ee
where $P=\int_0^\infty\d f S_a(f)$ is the total power of the local process.
The frequency $f_{\rm br}$ is a model parameter and has to be provided by the physical model of the initial fluctuations in the accretion disc. The breaking frequency is likely to be within the range between viscous $f_{\rm v}$ and Keplerian frequency $f_{\rm K}$ \citep{2015arXiv151205350H}: $f_{\rm v}<f_{\rm br}<f_{\rm K}$, which are separated by about 2 orders of magnitude (see Eq.\,(\ref{eq:frequencies})).
The integrated power of initial perturbations can depend on the radial coordinate in the accretion disc. \textit{In this paper, however, we take the integrated power to be constant over the accretion disc.}

The Lorentzian profile arises from the idea that the initial fluctuations in the time domain decay exponentially. Indeed, the Fourier transform of the function 
$g(t)=\sin(2\pi f_0 t)e^{-t/T}$, where $f_0$ and $T$ are constants, is given by 
$$
\overline{g}(f)\propto \frac{1}{((f-f_0)^2+(1/T)^2)^{1/2}}.
$$
The corresponding power is
$$
|\overline{g}(f)|^2\propto \frac{1}{(f-f_0)^2+(1/T)^2}.
$$
In that sense the Lorentzian profile given by equation (\ref{eq:LorentzProf}) implies that the initial fluctuations decay on timescale $\sim 1/\Delta f_{\rm br}$. In this paper we use zero-centured Lorentzian profiles, but if the accretion disc at some radial coordinate is disturbed by some process of characteristic frequency $f^*$, it would be reasonable to use the Lorentzian profile centured at frequency $f^*$.

\section{Fourier properties of the local mass accretion rate variability}

\subsection{Power spectrum}

The perturbations of the mass accretion rate originate from initial stochastic variability and are, therefore, also stochastic. They cannot be well described by any particular realisation of the mass accretion rate process in either time or frequency.  However, they can be described in terms of the average power spectrum and cross-spectrum. In the most general case when the mass accretion rate perturbations are given by (\ref{eq:mdot_tdomain}), PDS of the mass accretion rate at $R$ is defined by equation (see Appendix \ref{App:PSD}):
\beq
\label{eq:S_mdot}
S_{\dot{m}}(R,f)=\int\limits_{R_{\rm in}}^{R_{\rm out}}\frac{\d R'}{(R')^2}\Delta R(R')
|\overline{G}_{\dot{M}}(R,R',f)|^2 \nonumber \\
\times 
S_{a}(R',f)\otimes_f \left[\delta(f) + \frac{S_{\dot{m}}(R',f)}{\dot{M}^2_0}\right],
\eeq
where $S_{a}(R,f)$ is the power spectrum of initial perturbations at radius $R$, $\Delta R(R)$ is the radial scale on which the initial perturbations can be considered as coherent with one another. This scale can be different at different radii in the accretion disc. It is likely that this coherence scale is $\Delta R\sim H$, where $H$ is the geometrical thickness of the accretion disc at given radius \citep{2015arXiv151205350H}. However, in the case of accretion onto highly magnetized objects the scale can be larger.

\begin{figure}
\centering 
\includegraphics[width=8.5cm]{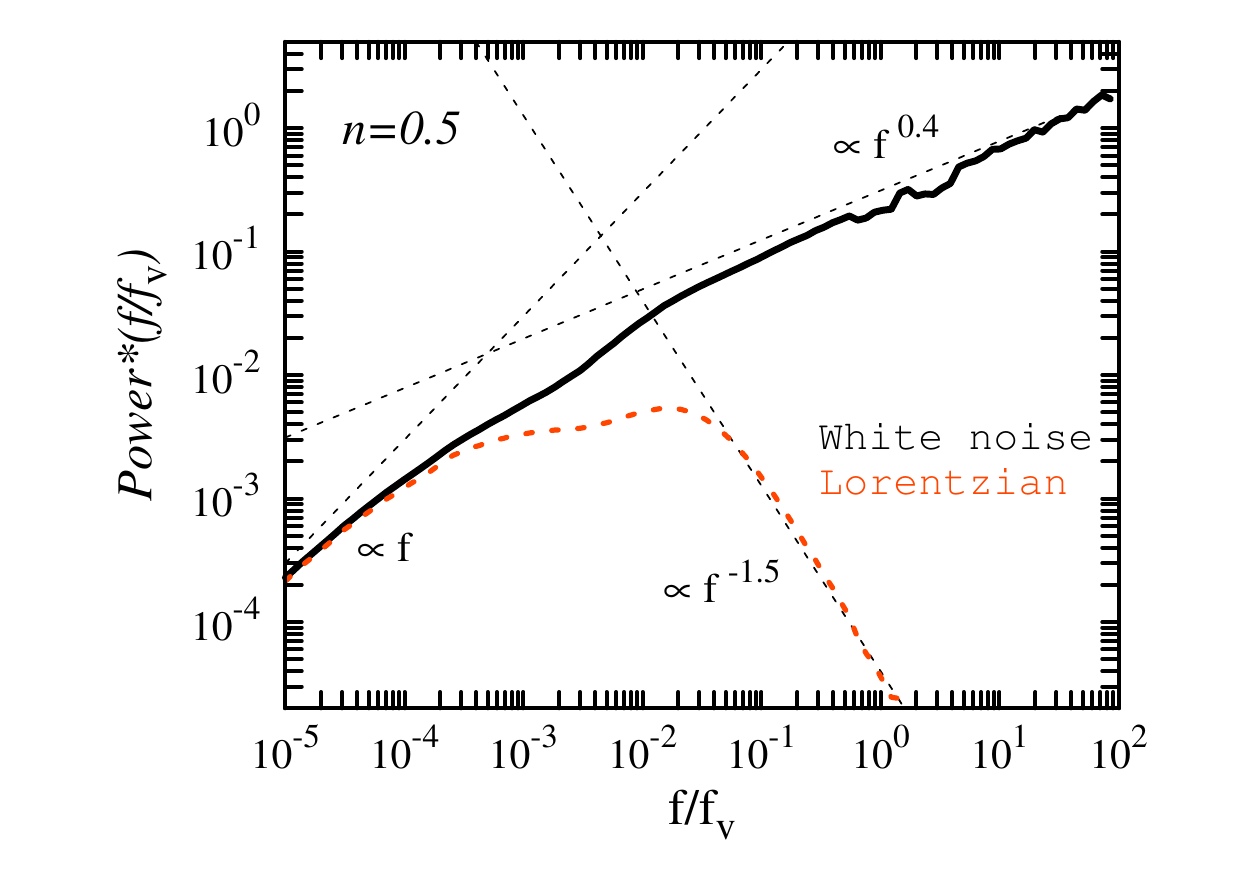} 
\includegraphics[width=8.5cm]{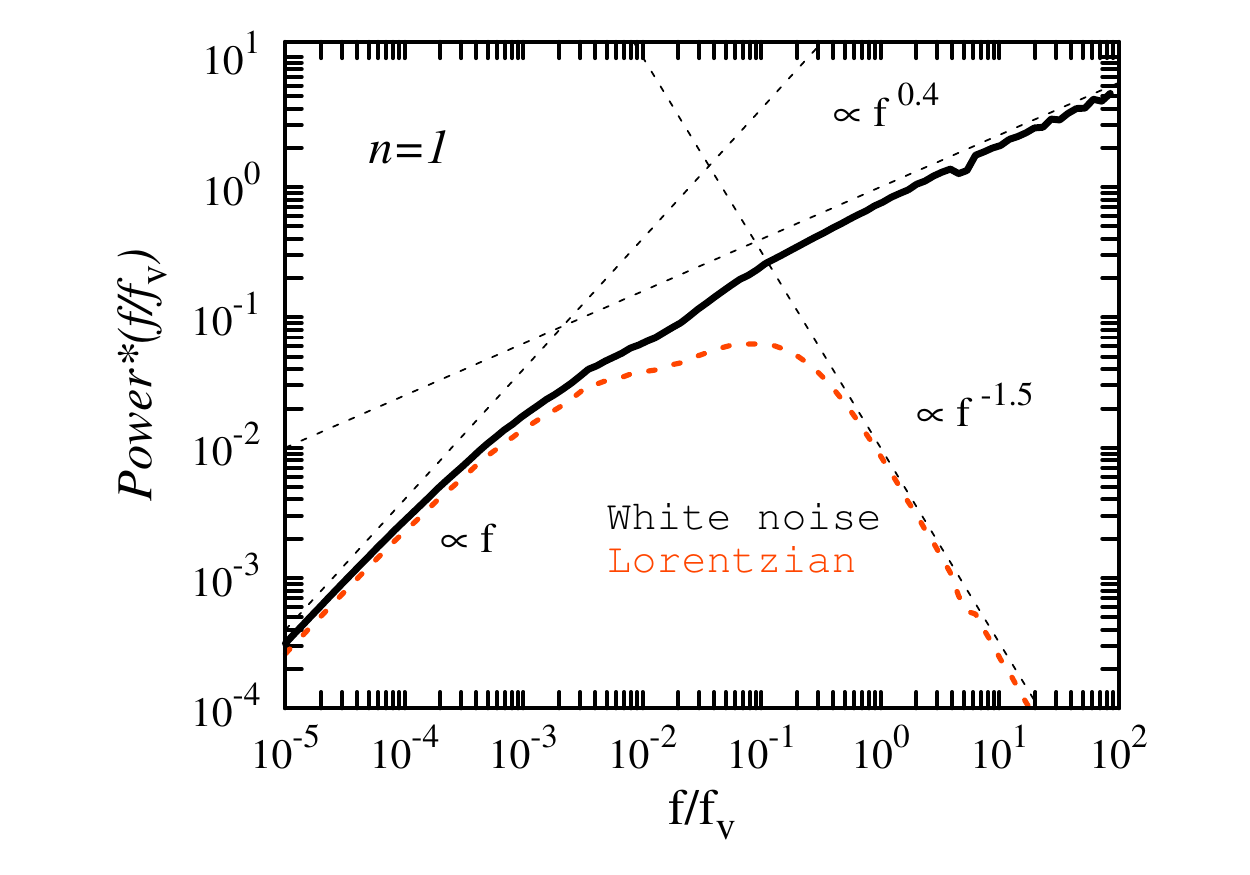}
\includegraphics[width=8.5cm]{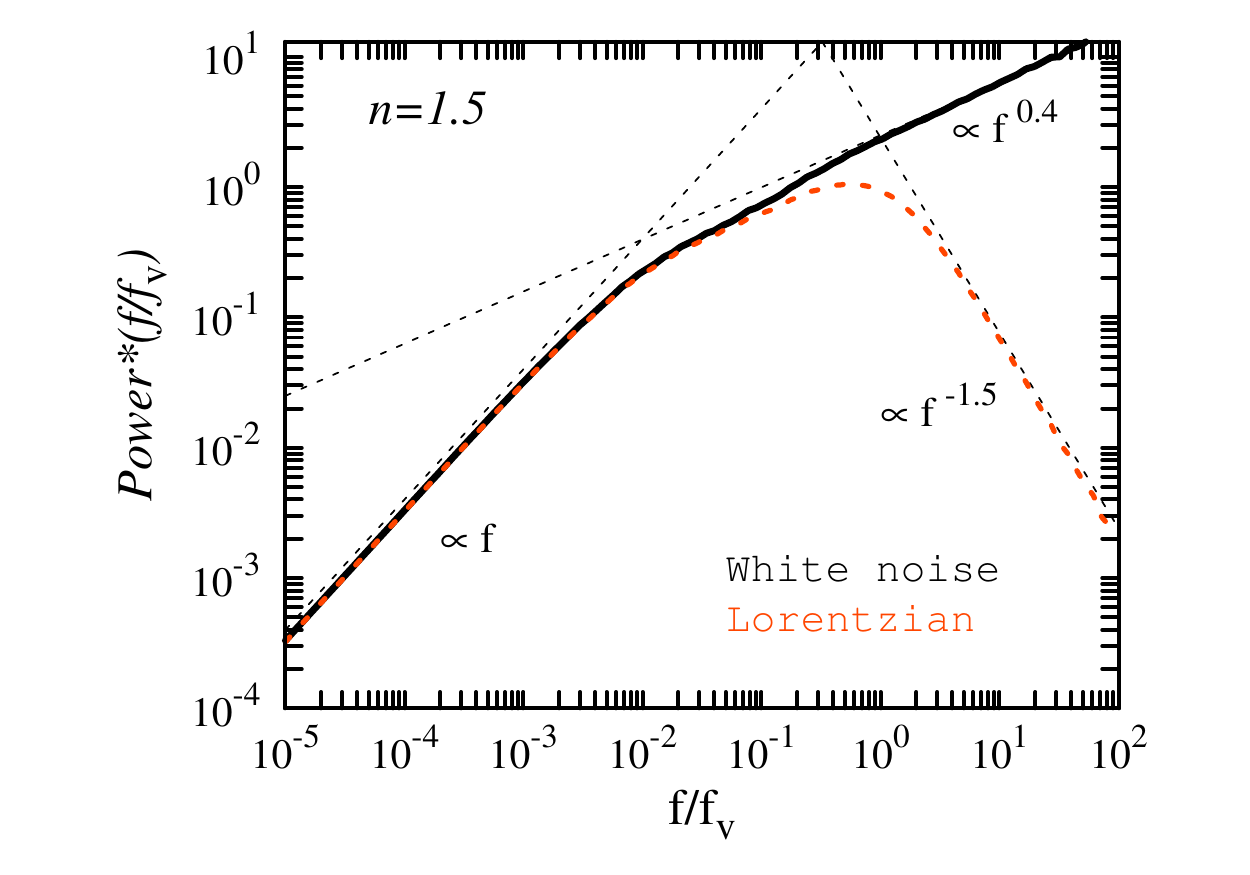}
\caption{The power spectra of the mass accretion rate at dimensionless radius $R=50$ for different parameters $n$. Black solid (red dashed) line corresponds to the initial fluctuations given by white noise (zero-centred Lorentzian breaking at local viscous frequency). The accretion disc inner radius is at $R_{\rm in}=0$. The initial fluctuations are produced at radii $R'\in(0,200)$. $\Delta R=1$.}
\label{pic:17_2}
\end{figure}

In the linear approach, when only the first term in the square brackets in the rhs of equation (\ref{eq:S_mdot}) is taken into account, the final power spectrum of the mass accretion rate is defined by the initial variability at every radius given by $S_{a}(R',f)$, and the transfer function power $|\overline{G}(R,R',f)|^2$. Because the transfer function suppresses variability at frequencies higher than local viscous frequency $f_{\rm v}(R')$, every region in the accretion disc effectively produces fluctuations at frequencies below the corresponding $f_{\rm v}(R')$.  As a result, there should be a break in the power spectrum, which corresponds to the lowest viscous frequency in that part of the accretion disc, where initial fluctuations are produced. The lowest frequency corresponds to the outer radius $R_{\rm out}$ and the break arises even in the case when the initial perturbations are given by white noise (see Fig.\,\ref{pic:17_2}). If the initial fluctuations are produced below some local frequency $f_{\rm br}(R)$ only, then there should be another break in power spectrum, which locates at $\sim f_{\rm br}(R)$ (see red dashed line in Fig.\,\ref{pic:17_2}). At different radii the location of the second break is also different in general. Above this frequency the variability is not produced effectively by local processes and additionally suppressed by the transfer function. Thus, one would expect double broken power spectrum with breaks corresponding to the inner and outer disc radii.

The inclination of the power spectrum between two slopes and above the second slope depends on the power spectra of initial perturbations. We describe the power spectrum of initial perturbations by zero-centred Lorentzians breaking at frequency $f_{\rm br}(R)$. If, for example, the power spectrum drops exponentially at frequencies above $f_{\rm br}(R)$ (like in the model of \citealt{1997MNRAS.292..679L}), the final power spectrum is steeper. In this case it is, indeed, $\propto 1/f$ between two slopes as found by \citealt{1997MNRAS.292..679L}.

The second term in equation (\ref{eq:S_mdot}) describes the response of the newly arising fluctuations on local variability of the mass accretion rate (the amplitude of new fluctuations is proportional to the local mass accretion rate, which is variable). This term includes all non-linear interactions between inward, outward and local fluctuations. The second term is negligible if the mass accretion rate variability amplitude is much smaller than the average mass accretion rate.

Because the initial fluctuations of the surface density scale with the mass accretion rate, the PDS of initial perturbations is $S_{a}(R,f)\propto \dot{M}^2$. 
The rms of local fluctuations of the mass accretion rate is
$$({\rm rms})^2 \propto \int\limits_{f>0} \d f\, S_{\dot{m}}(f)$$ 
Therefore, if the nonlinear effects are not taken into account and there is only the first term in the integrand in equation (\ref{eq:S_mdot}), the rms of local mass accretion rate is proportional to the mass accretion rate:
\be 
{\rm rms}\propto \dot{M}.
\ee
This provides a neat explanation for the linear rms-flux relation ubiquitously observed from accreting objects.

Interestingly, the nonlinear term in equation (\ref{eq:S_mdot}) can cause a deviation from this linear rms-flux relation.

\subsection{Cross-spectrum}

The cross-spectrum of mass accretion rate variability for radii $R_1$ and $R_2$ is (see Appendix \ref{App:Cross})
\beq
\label{eq:C_mdot}
C_{\dot{m}}(R_1,R_2\,|\,f)&=\int\limits_{R_{\rm in}}^{R_{\rm out}}\frac{\d R'}{(R')^2}\Delta R(R')\overline{G}_{\dot{M}}(R_1,R',f)\overline{G}^*_{\dot{M}}(R_2,R',f)& \nonumber \\
& \times S_{a}(R',f)\otimes_f
\left[\delta(f)+\frac{S_{\dot{m}}(R',f)}{\dot{M}^2_0}\right],&
\eeq
where $S_{a}(R,f)$ is the PDS of initial perturbations, $S_{\dot{m}}(R,f)$ is PDS of the mass accretion rate defined by (\ref{eq:S_mdot}), $\Delta R$ is local radial scale, on which the initial fluctuations are correlated, and `*' denotes complex conjugation. The second term in the rhs of (\ref{eq:C_mdot}) is negligible if the perturbations of the mass accretion rate are much smaller than the average mass accretion rate. It is obvious that $C_{\dot{m}}(R_2,R_1\,|\,f)=C^*_{\dot{m}}(R_1,R_2\,|\,f)$.
It is also obvious that the cross-spectrum gives the power spectrum when $R_1=R_2$: $C_{\dot{m}}(R,R\,|\,f)=S_{\dot{m}}(R,f)$.

In order to measure the coherence between two radii $R_1$ and $R_2$ it is convenient to use coherence function:
\be
{\rm Coh}(R_1,R_2\,|\,f) \equiv\frac{|C_{\dot{m}}(R_1,R_2\,|\,f)|^2}{S_{\dot{m}}(R_1,f)S_{\dot{m}}(R_2,f)},
\ee
which takes the value within $[0,1]$.

\begin{figure}
\centering 
\includegraphics[width=8.5cm]{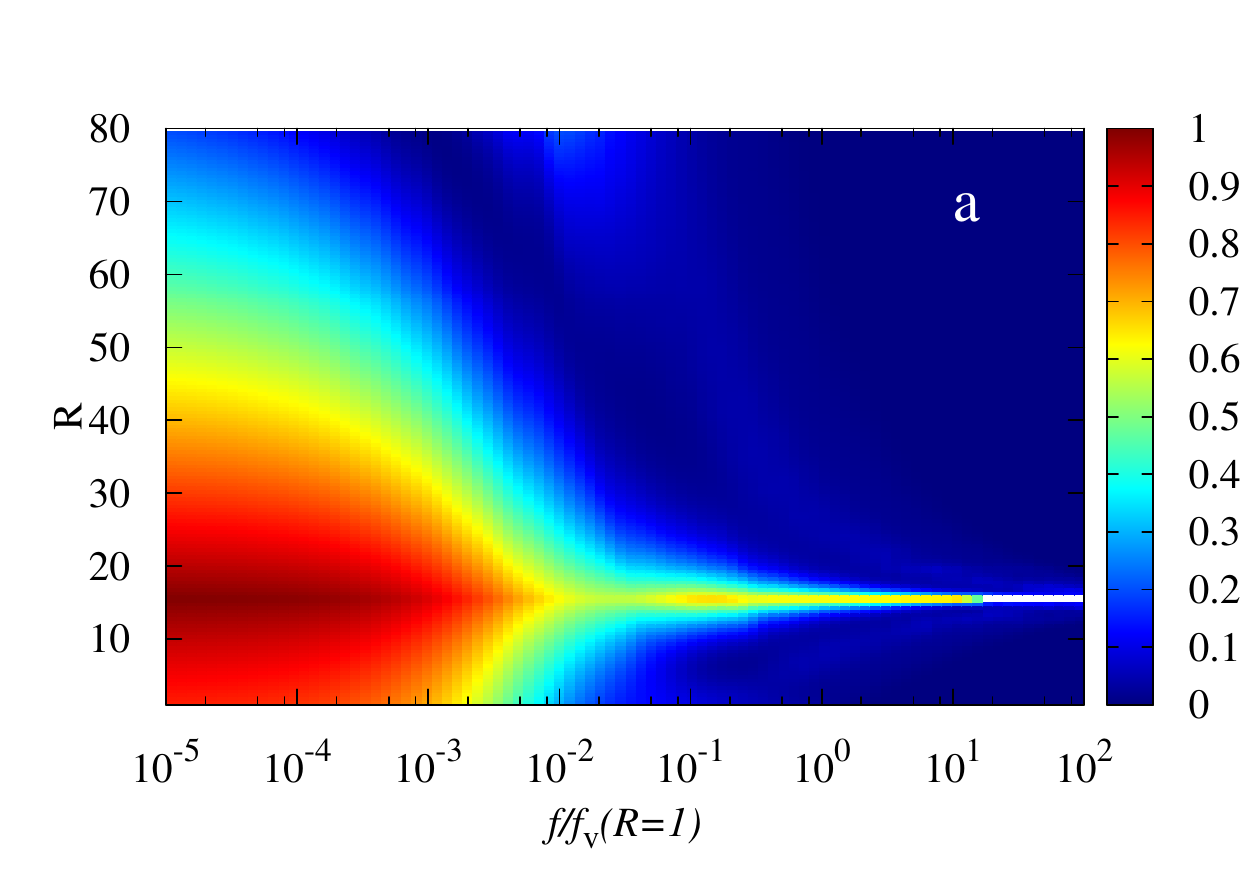} 
\includegraphics[width=8.5cm]{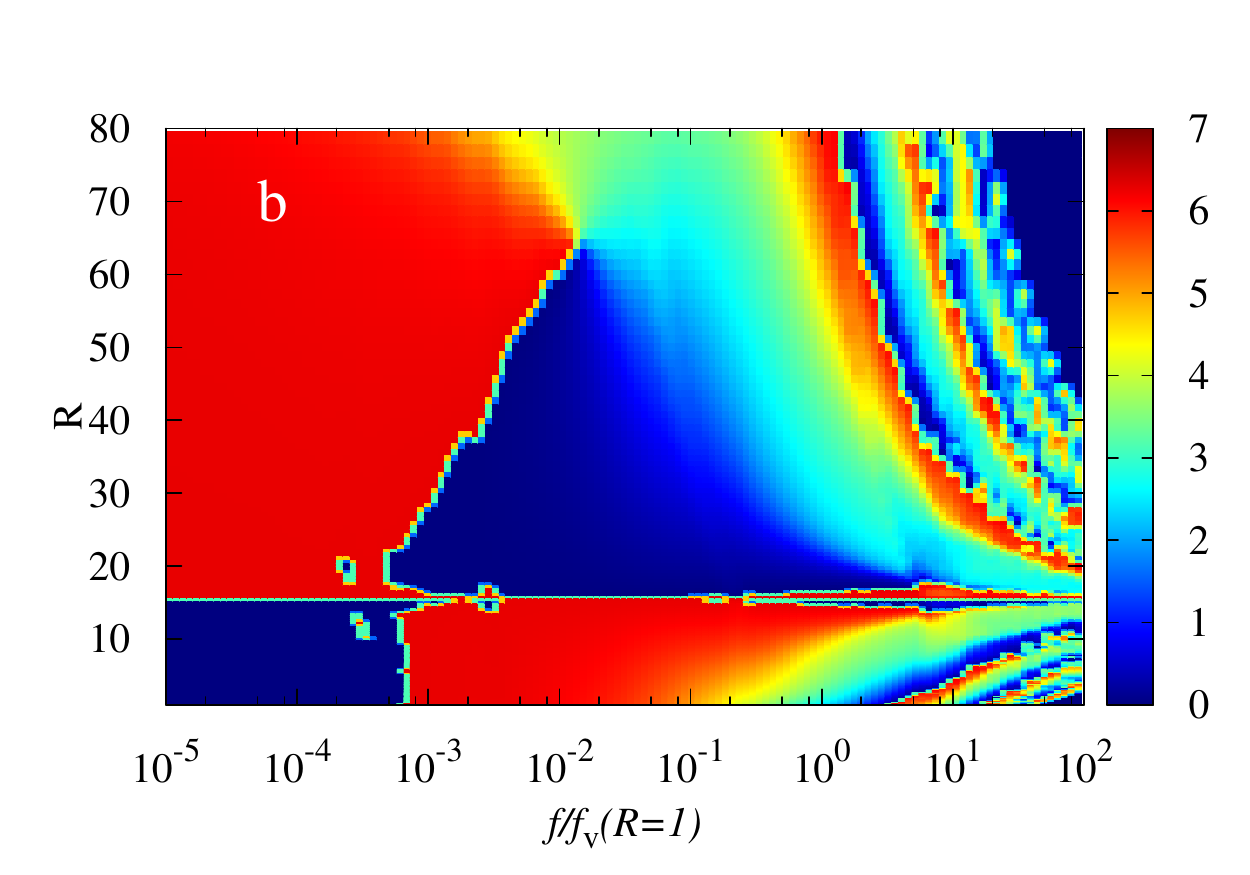} 
\caption{{(a) The coherence function ${\rm Coh}(15,R\,|\,f)$ for variability at radial coordinate $R_1=15$ and other radii in the accretion disc. (b) The corresponding phase angle of the cross-spectrum. At low Fourier frequencies the variability at outer radii leads variability at $R=15$ (red color) and variability at inner radii lags variability at $R=15$. At sufficiently high frequencies the situation is opposite: the variability at smaller radii lead variability at larger radii. Parameters: $n=1$, $R_{\rm in}=0$, $R_{\rm out}=100$, $\Delta R=1$, $f_{\rm br}=100 f_{\rm v}$.}}
\label{pic:23}
\end{figure}

Because the high frequency variability is suppressed by the transfer functions, coherent high frequency variability is taking place only at radii close to the origin of the initial perturbations (see Fig.\,\ref{pic:23}a). At low frequencies the variability is coherent at a broader range of radii. It is interesting that at low frequencies the mass accretion rate variability at smaller radii lags variability at larger radii, however, at high frequency (see Fig.\,\ref{pic:23}b) the situation might be opposite: the variability at larger radii lags relative to the variability at smaller radii. This happens for two reasons: (i) the mass accretion rate fluctuations can propagate outwards and (ii) the viscous frequency drops with radial coordinate and at sufficiently high frequency, variability from outer parts of a disc is more suppressed then variability from inner parts of accretion disc, i.e. fluctuations from outside are damped out whereas those propagating outwards from smaller radii are not. 

\section{Fourier properties of the output flux} 

In this section we discuss the fourier properties of the output flux mainly in context of BHs, for which the flux is coming entirely from the accretion disc. 
We assume that the total flux at the surface of the accretion disc drops with the radial coordinate as $F_{\rm sur}\propto R^{-3}$.

\subsection{Power spectra for a given energy band}

The total luminosity available to be radiated at a given region of the accretion disc is proportional to the local mass accretion rate. If local variability of the mass accretion rate is small in comparison with the average mass accretion rate, the fluctuations of flux in some energy band are $\propto \dot{m}(R,t)$. If the energy dependence stays constant in time, the flux in a given energy band is
\be\label{eq:Flux}
F_{\rm h}(t)\simeq\int\limits_{R_{\rm in}}^{R_{\rm out}}\d R'\,\frac{h(R')}{R'^2}\dot{m}(R',t),
\ee
where $h(R)$ is a weighting function which corresponds to the energy band. Then the PDS of the flux variability in the energy band is
\be
S_{F_{\rm h}}(f)=\int \limits_{R_{\rm in}}^{R_{\rm out}}\d R_1\,\int \limits_{R_{\rm in}}^{R_{\rm out}}\d R_2\, \frac{h(R_1)}{R_1^2}\frac{h(R_2)}{R_2^2} C_{\dot{m}}(R_1,R_2\,|\,f),
\ee
where $S_{\dot{m}}(R,f)$ is PDS of local mass accretion rate given by equation (\ref{eq:S_mdot}).

For simplicity the weighting function $h(R)$ for the hard energy band is taken to be either a step function:
\begin {equation} 
h(R) \! =\!  
\left\{  \begin{array}{ll}
\strut\displaystyle \! \! 
1, & \mbox{for}\ R<3R_{\rm in}, \\
\! \!  \frac{1}{3},                                             & \mbox{for}\ R\geq 3 R_{\rm in}, 
 \end{array} \right. \nonumber
\end{equation} 
{which corresponds to abrupt hardening of the spectrum within the innermost region or exponent: $h(R)\propto R^{-1}$, which corresponds to gradual hardening of the spectrum}. The weighting function for the soft energy band is taken to be $s(R)=1$ all over the accretion disc.

\begin{figure}
\centering 
\includegraphics[width=8.7cm]{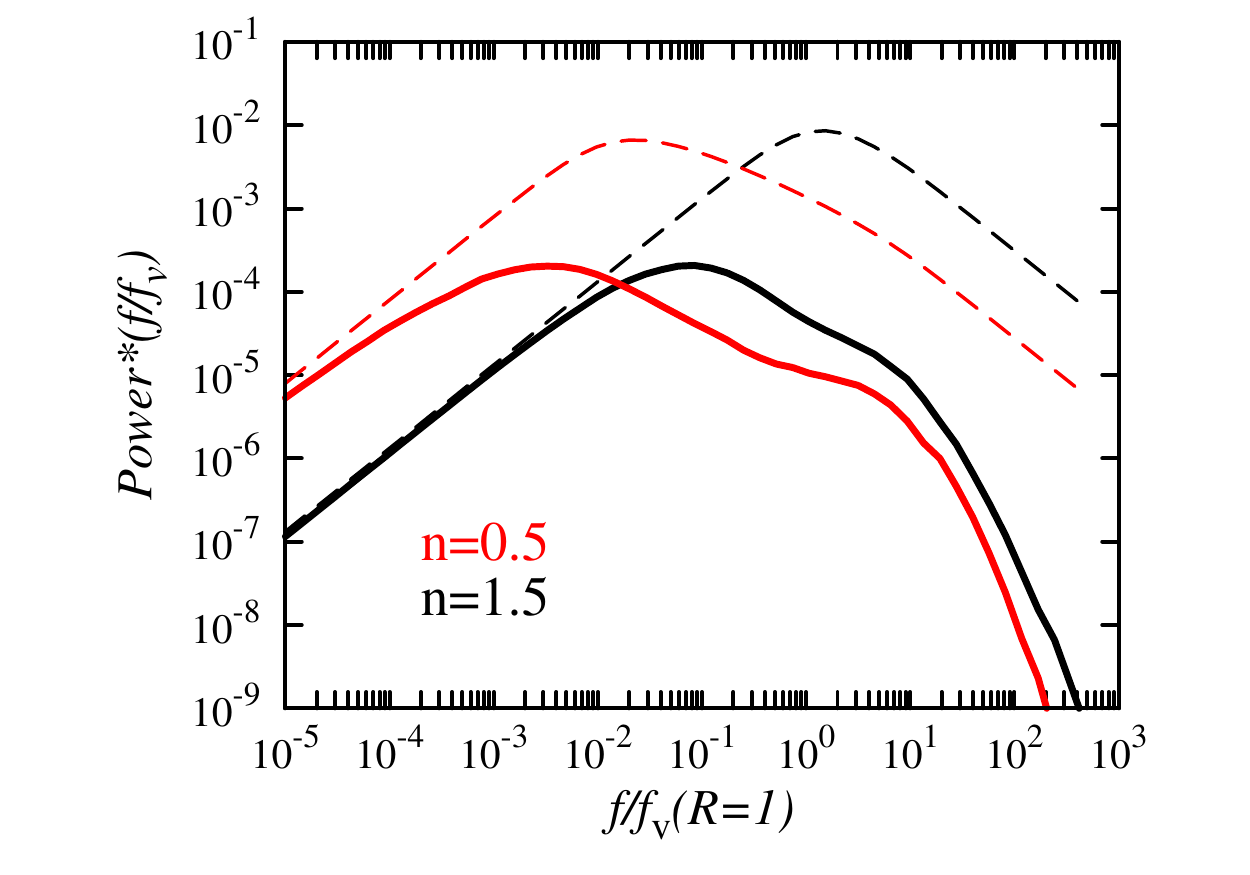} 
\caption{{The power spectrum of total flux variability. The solid lines represent the result based on accurate Linden-Bell Green functions, while the dashed lines show the result based on Green functions given by $\delta$-function in the time domain. The real diffusion process suppress significantly the variability at high frequencies. Parameters: $R_{\rm in}=1$, $R_{\rm out}=100$, rms$_{\rm ini}=0.01$. The initial mass accretion rate perturbations are described by zero-centured Lorentzian profile with $f_{\rm br}=f_{\rm v}$. The hardness function is given by the step function.}}
\label{pic:PowerFlux} 
\end{figure}

\begin{figure}
\centering 
\includegraphics[width=8.7cm]{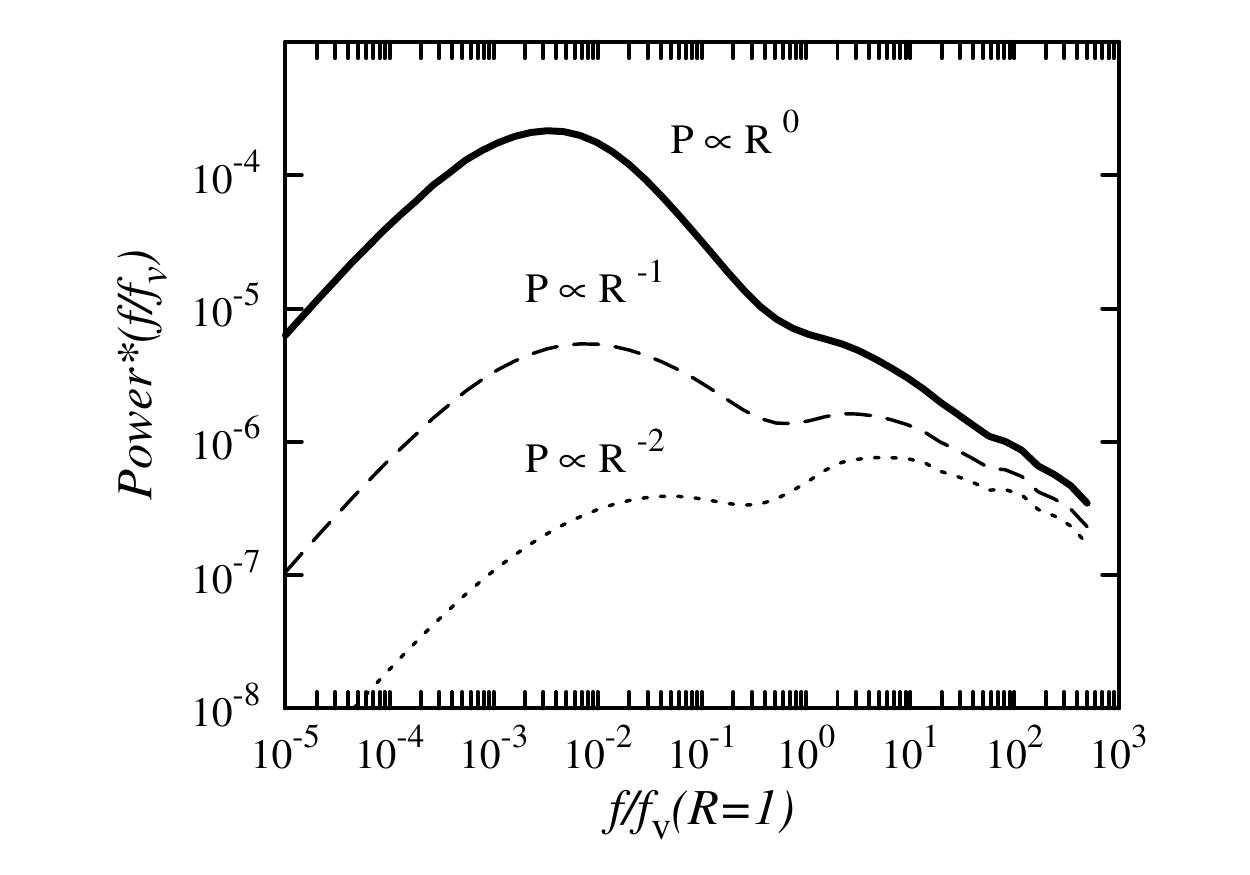} 
\caption{{The power spectra of total flux variability given for different distribution of power of initial variability over the accretion disc. The initial mass accretion rate perturbations are described by a zero-centured Lorentzian profile with $f_{\rm br}=100\,f_{\rm v}$. Parameters: $R_{\rm in}=1$, $R_{\rm out}=100$, $\Delta R=1$. The hardness function is given by the step function.}}
\label{pic:PowerFlux2} 
\end{figure}

The power spectrum of the photon flux variability imprints information about the inner and outer radii of the accretion disc and, like the power spectrum of the mass accretion rate, has at least two breaks at about the viscous frequencies at inner and outer radii (see Fig.\,\ref{pic:PowerFlux}). The shape of the power spectrum strongly depends on the distribution of power of initial mass accretion rate fluctuations over the radial coordinate (see Fig.\,\ref{pic:PowerFlux2}).

\begin{figure}
\centering 
\includegraphics[width=9cm]{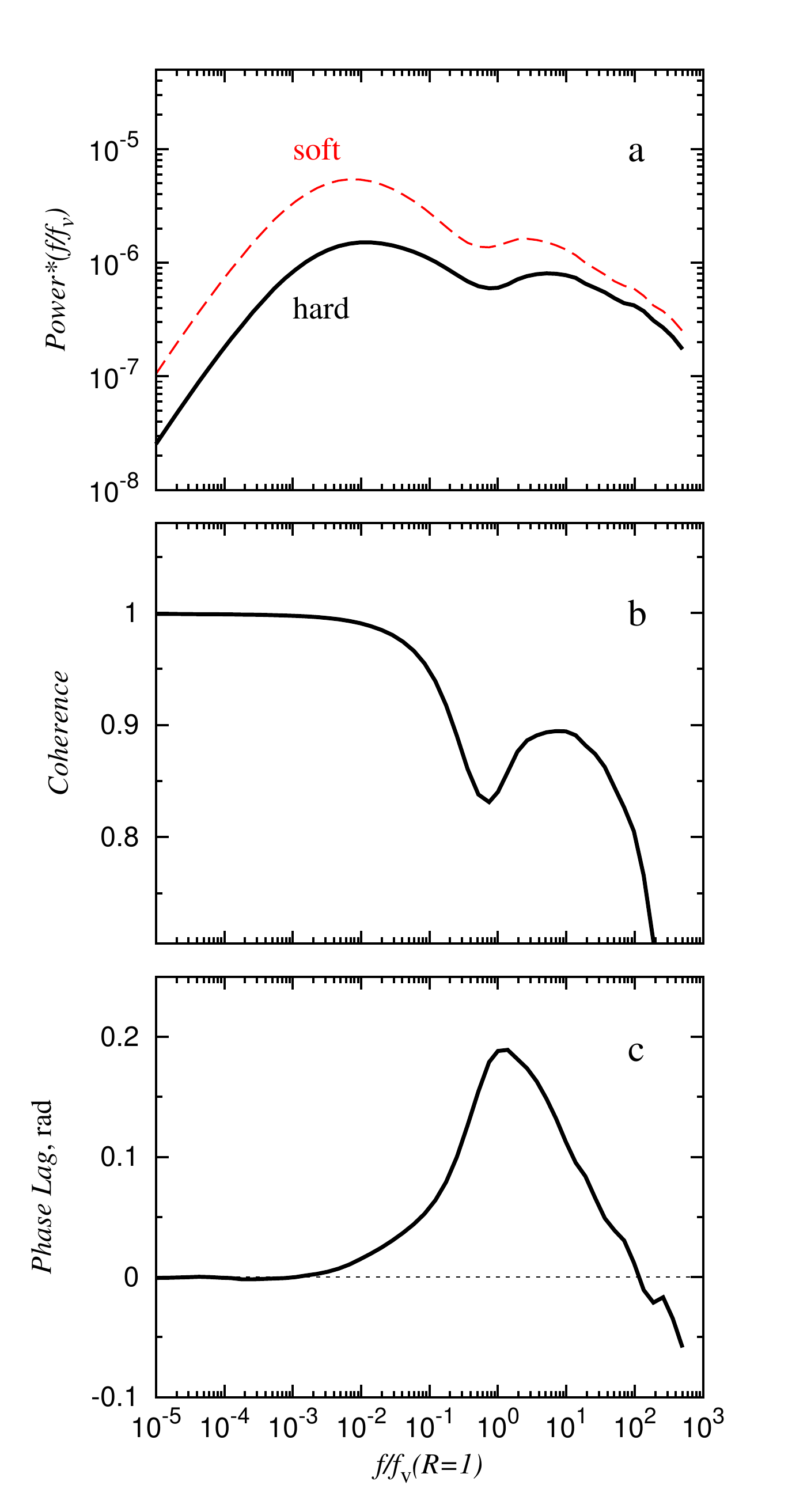} 
\caption{{(a) Power spectra of variability in hard (solid black line) and soft energy bands (dashed red line), (b) the coherence function between variability at hard and soft energy bands and (c) the phase lags between variability in hard and soft energy bands. The initial mass accretion rate perturbations are described by zero-centured Lorentzian profile with $f_{\rm br}=100\,f_{\rm v}$. Parameters: $R_{\rm in}=1$, $R_{\rm out}=100$, $\Delta R=1$. The hardness function is taken to be $\propto R^{-1}$.}}
\label{pic:PowerFlux3} 
\end{figure}

\subsection{Cross-spectrum between two energy bands: hard and soft lags}

One can consider two energy bands with different emissivities given by weighting functions $h(R)$ (hard band) and $s(R)$ (soft band), which result in different power spectra in hard and soft energy bands (see Fig.\,\ref{pic:PowerFlux3}a). Then the cross-spectrum is calculated as follows: 
\be\label{eq:CrossEnergyBand}
C_{\rm h,s}(f)= \int \limits_{R_{\rm in}}^{R_{\rm out}}\d R_1 \int \limits_{R_{\rm in}}^{R_{\rm out}}\d R_2\,
\frac{h(R_1)}{R^2_1} \frac{s(R_2)}{R^2_2} C_{\dot{m}}(R_1,R_2\,|\,f),
\ee
where $C_{\dot{m}}(R_1,R_2\,|\,f)$ is the cross-spectrum for the mass accretion rate given by equation (\ref{eq:C_mdot}). 
{The corresponding coherence function (see Fig.\,\ref{pic:PowerFlux3}b) is given by
\be
{\rm Coh}_{\rm h,s}(f)=\frac{\left| C_{\rm h,s}(f)  \right|^2}{S_{F_{\rm h}}(f) S_{F_{\rm s}}(f)}, 
\ee
}
while the phase lag $\Phi(f)$ between the two energy bands (see Fig.\,\ref{pic:PowerFlux3}c) is given by
\be
\tan\Phi (f)=\frac{{\rm Im}[C_{\rm h,s}(f)]}{{\rm Re}[C_{\rm h,s}(f)]},
\ee
where ${\rm Im}[C_{\rm h,s}(f)]$ and ${\rm Re}[C_{\rm h,s}(f)]$ are imaginary and real parts of the cross-spectrum at frequency $f$. The corresponding time lag is
$t_{\rm lag}(f)=\Phi (f)/(2\pi f)$.

\begin{figure}
\centering 
\includegraphics[width=8.7cm]{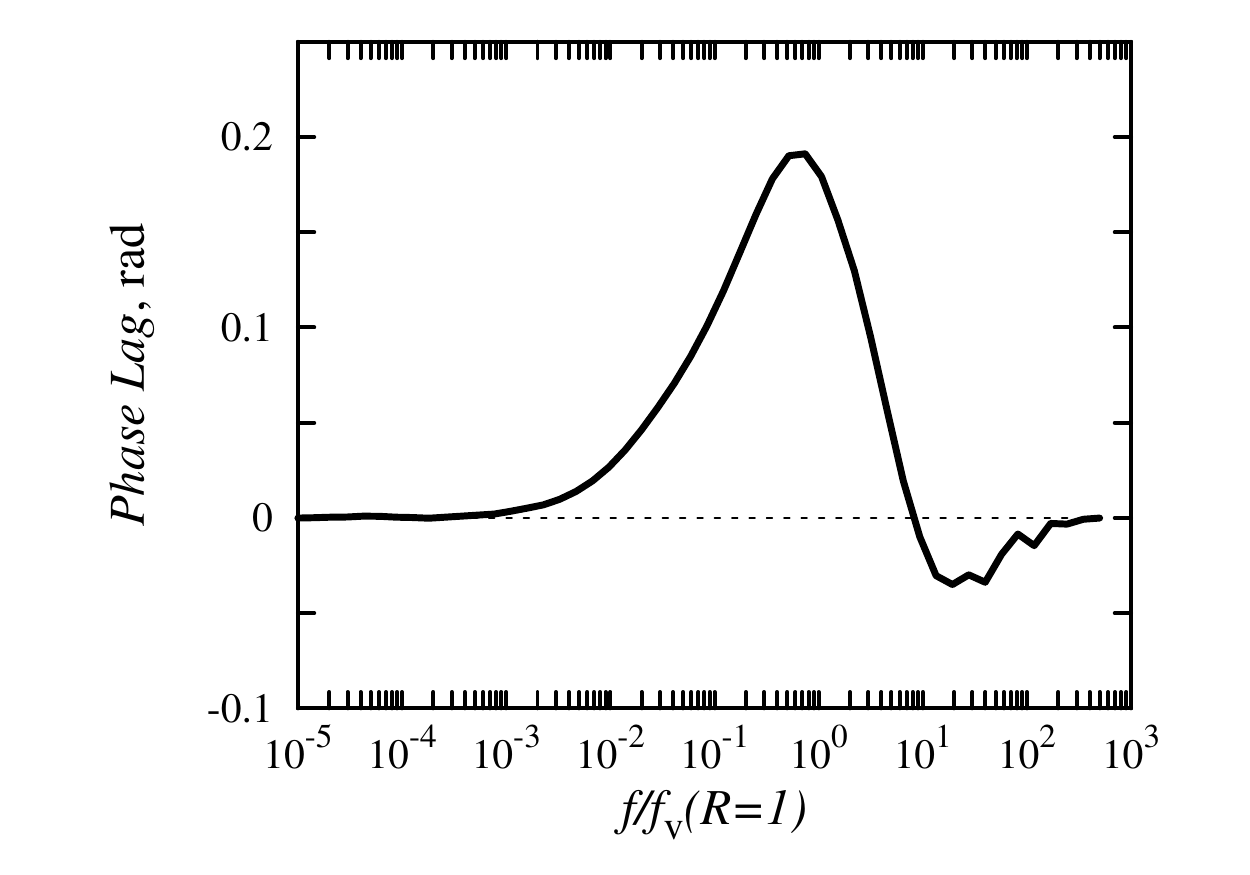} 
\caption{The phase lags as a function of Fourier frequency. Hard X-ray variability lags soft X-ray variability at low frequencies. However, high frequencies soft variability lags hard variability. It is caused by principal possibility of diffusion from inner to outer radii. Parameters: $R_{\rm in}=1$, $R_{\rm out}=100$. The initial perturbations are given by zero-centured Lorentzian profile. The hardness function is given by step function.}
\label{pic:PhaseLag} 
\end{figure}

It is natural that the variability in the hard energy band lags the variability in the soft energy band. In this case the time lags are defined by the typical time of propagation of fluctuations from outer radii, where soft flux is produced, to inner radii, where the most energy flux is produced in the hard energy band. However, negative phase lags (when the soft energy band variability lags hard energy variability) are also possible at high frequencies (see Fig.\,\ref{pic:PhaseLag}). The negative lags are caused by the fact that the variability of mass accretion rate at the inner radii can affect the variability at the outer radii. It becomes important at high frequencies where the Green functions in the frequency domain suppress variability from the outer radii. The frequency where the soft variability lags hard variability is comparable to the viscous frequency at the inner disc radius. {It is important to note that the appearance of the negative phase lags is strongly dependent on specific conditions in the accretion disc. Moreover, negative lags arise at very high Fourier frequencies, where the variability can be principally affected by sound waves instead of viscous diffusion.}

The predicted negative phase lags are comparable to the observed negative lags in stellar mass black hole systems \citep{2011MNRAS.414L..60U,2015ApJ...814...50D} and AGN \citep{2010MNRAS.401.2419Z,2013ApJ...777L..23W,2014MNRAS.439.1548A}. Therefore, viscous diffusion in accretion disc can in principle be a key process responsible for negative time lags, but we note that Fe K lags in AGNs are solid evidence of reverberation.

The coherence is typically smaller at higher Fourier frequency. It is a result of suppression of mass accretion rate variability due to viscous diffusion, when the variability at high frequency can be coherent only at close radii. However, because of the possibility of outward propagation (note that outward propagation effectively transfers variability of a particular characteristic frequency, suppressing variability at higher and lower frequencies), there is a range of Fourier frequencies, for which the coherence becomes higher under the additional influence of outward propagation (see the high frequency bump in Fig.\,\ref{pic:PowerFlux3}b). Depending on the displacement of regions producing hard and soft photon energy flux, the coherence becomes higher or even reaches its local maximum at sufficiently high frequency. It is remarkable, that the increasing of coherence at high frequencies is expected also from simple reverberation models \citep{2010MNRAS.401.2419Z,2014A&ARv..22...72U}.

\subsection{Influence of non-linearity}

\begin{figure}
\centering 
\includegraphics[width=8.7cm]{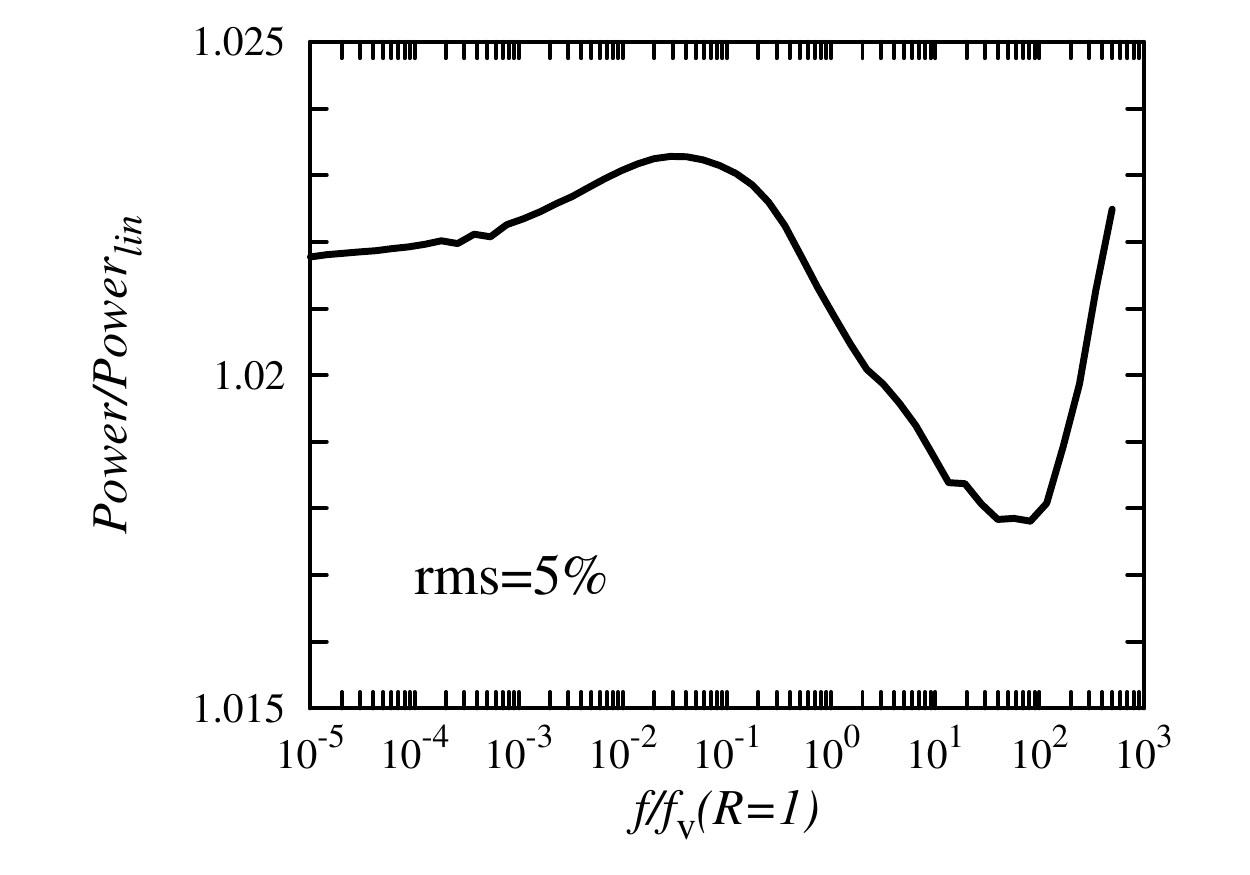} 
\caption{{The ratio of PDS with non-linear effects taken into account to PDS without non-linearity. Non-linear effects increase PDS.
 The initial mass accretion rate perturbations are described by zero-centured Lorentzian profile with $f_{\rm br}=100\,f_{\rm v}$. Parameters: $R_{\rm in}=1$, $R_{\rm out}=100$, $\Delta R=1$. The hardness function is taken to be $\propto R^{-1}$.}}
\label{pic:NonLin1} 
\end{figure}

\begin{figure}
\centering 
\includegraphics[width=8.7cm]{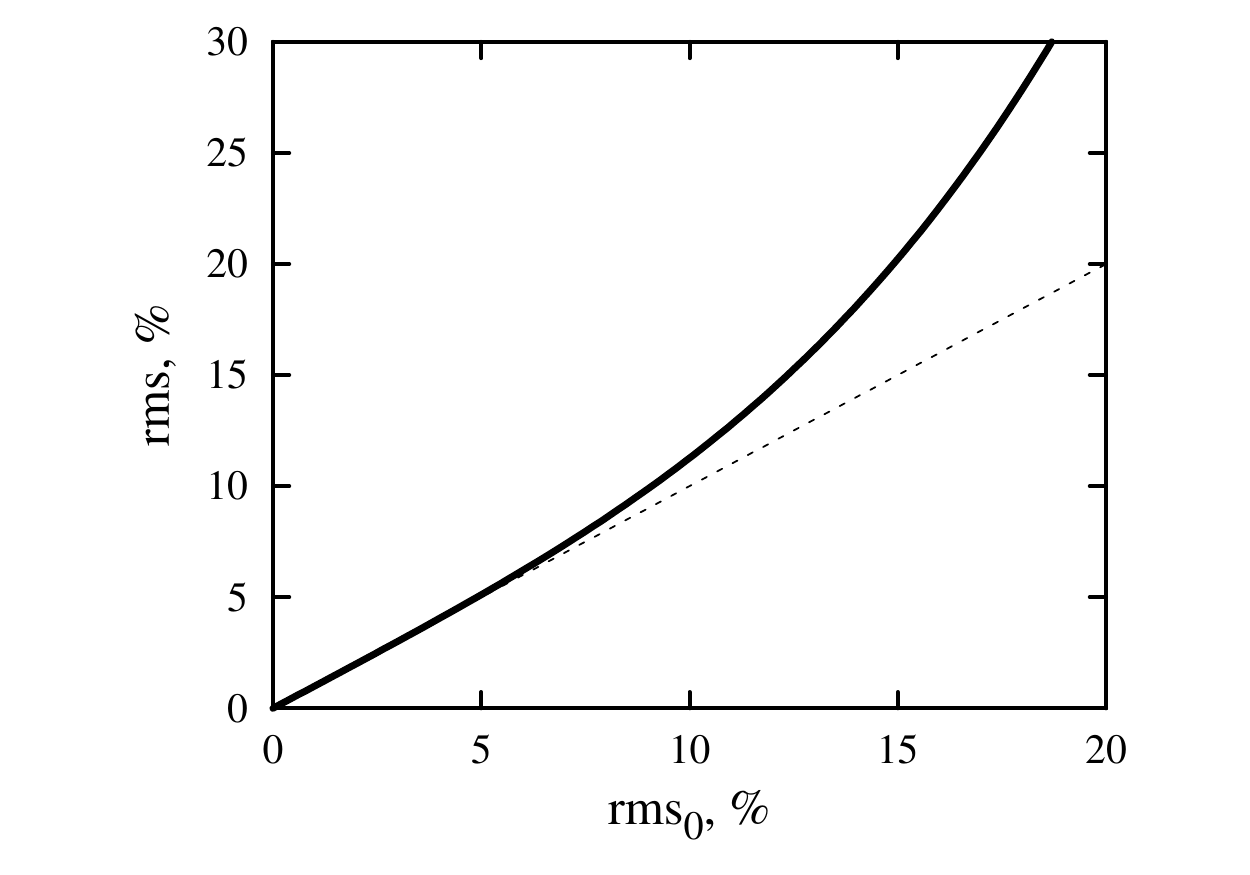} 
\caption{{The rms of total flux variability with effects of non-linearity taken into account as a function of rms, where non-linear effects are neglected (rms$_{0}$, dashed line represents rms=rms$_{0}$ corresponding to linear case). The stronger the variability, the stronger non-linear effects. The initial mass accretion rate perturbations are described by zero-centured Lorentzian profile with $f_{\rm br}=100\,f_{\rm v}$. Parameters: $R_{\rm in}=1$, $R_{\rm out}=100$, $\Delta R=1$. The hardness function is taken to be $\propto R^{-1}$.}}
\label{pic:NonLin2} 
\end{figure}

New local fluctuations of the surface density arise on top of the average mass accretion rate at each radial coordinate . Thus, new fluctuations can be affected by existing variability in the accretion flow, which gives rise to non-linear effects. The non-linear effects are described by the second terms in the rhs of equation\,(\ref{eq:S_mdot}) and equation\,(\ref{eq:C_mdot}).
The detailed study of non-linear effects is beyond the scope of the present work. However, a few features have to be marked: (i) the contribution of non-linear effects influence the shape of the PDS shape and Fourier frequency (see Fig.\,\ref{pic:NonLin1}), (ii) the higher the relative rms, the stronger the non-linear effects (see Fig.\,\ref{pic:NonLin2}).

\subsection{Accreting neutron stars}

In the case of highly magnetized accreting NSs the accretion disc is interrupted at the magnetospheric radius, where the accretion disc temperature is too low to produce X-ray flux (unless the mass accretion rate is extremely high, as it is for the case of pulsating ultra luminous X-ray sources powered by NSs). Most of the flux is produced at the NS surface, where the matter channeled by the magnetic field looses its kinetic energy (however, at extremely high mass accretion rate the flux can be affected by the matter at the magnetosphere, see e.g. \citealt{2017MNRAS.tmp..143M}). As a result, the observed fast aperiodic variability of the X-ray flux in the case of highly magnetized accreting NS is caused by mass accretion rate variability at the thin inner ring of the accretion disc. This feature causes observed variability of the PDS with the luminosity: the higher the mass accretion rate, the smaller the inner disc radius, the higher the Fourier frequency of a break corresponding to the inner disc radius \citep{2009A&A...507.1211R}. The PDS of observed X-ray variability, therefore, imprints information about the distant regions of accretion disc \citep{2005astro.ph..1215G}, which is useful for diagnostics of the whole system. It is interesting that accreting highly magnetized NSs provide an opportunity to study timing properties of cold accretion discs, where the temperature drops below $\sim 6500\,{\rm K}$ and hydrogen is weakly ionized and viscous properties of the disc can be significantly different \citep{2017arXiv170304528T}. The same condition of an accretion disc can be realized in BH XRBs, but only at much lower mass accretion rates.

\section{Summary}
\label{discussion}

We have developed a formalism that describes propagation of mass accretion rate fluctuations in an accretion disc using an exact solution of the diffusion equation \citep{1974MNRAS.168..603L}. The solution of the equation is given by Green functions, which can be constructed for various boundary conditions. The Green functions in the frequency domain play the role of transfer functions and describe how initial variability at a given radius results in variability at other radii. The fluctuations of the surface density arising at some radius due to local processes result in mass accretion fluctuations propagating both inwards and outwards. The developed formalism provides the basis for the construction of accurate models of fast aperiodic variability of mass accretion rate in accretion discs around BHs and NSs. 

It is shown that the Green functions suppress variability at frequencies higher than the local viscous frequency due to diffusive spreading. The suppression depends on the Fourier frequency of variability and is stronger for lower parameter $n$, which describes dependence of kinematic viscosity on the radial coordinate $\nu\propto R^n$ (see Fig.\,\ref{pic:PS13}a and Fig.\,\ref{pic:PS18}a). As a result, the outer parts of the accretion disc effectively produce variability only below their viscous frequency even if high frequency variability is produced locally. High frequency variability can propagate only to the neighboring radii without significant suppression. The suppression at high Fourier frequencies encoded in Green functions is in agreement with recent numerical simulations based on the first principles \citep{2014ApJ...791..126C,2015arXiv151205350H}. This property of viscous diffusion means that different parts of the accretion disc effectively produce mass accretion rate fluctuations only at timescales longer than the local viscous time scale.

It is interesting that according to solutions of the diffusion equation, the mass accretion rate fluctuations propagate outwards as well, which has not been explored before in the context of propagating mass accretion rate fluctuations. In these conditions, the outer parts of accretion disc ``feel" processes which occur in the inner part of the disc. However, in the case of outward propagation of fluctuations, the transfer functions (see Fig.\,\ref{pic:PS13}b and Fig.\,\ref{pic:PS18}b) differ from the transfer function describing inward propagation. The high frequency variability is strongly suppressed similarly to the case of inward propagation, but the transfer functions have a maximum at a frequency close to local viscous frequency. It means that the low frequency variability is also suppressed for fluctuations propagating outwards. In this case the suppression is stronger for lower frequencies and higher parameter $n$.

The power spectrum of the mass accretion rate variability at any radial coordinate is affected by properties of initial fluctuations all over the disc (distribution of variability power over the radial coordinate and break frequency $f_{\rm br}(R)$ of initial variability) and by viscous properties of the accretion disc, which are described by Green functions. The observed variability of the photon energy flux in a given energy band is defined by mass accretion rate variability at the inner parts of accretion disc. Thus, it is in principal possible to decode the properties of initial variability using observed photon energy flux variability and appropriate Green functions. 

Our model assumes that there is an effective radial scale $\Delta R$ in the accretion disc where the initial fluctuations are correlated. This assumption is in agreement with recent numerical simulations \citep{2015arXiv151205350H}, where the radial scale is associated with the typical scale at which the dynamo process develops \citep{2004MNRAS.348..111K}. The influence of initial fluctuations on the final variability depends on the local radial scale $\Delta R$. {Previous, discretized, treatments of the propagating fluctuations model (e.g. \citealt{2013MNRAS.434.1476I}) also implicitly make this assumption.}

The developed model naturally predicts two breaks in the PDS: the lower frequency break corresponds to the outer disc radius and settles at about the viscous frequency of the outer radius, the higher frequency break corresponds to the inner disc radius, its location depends on viscous frequency at the inner radius and PDS of initial perturbation there. Between the breaks the PDS can be described by power laws except the highest frequencies, where the PDS can drop faster than it is described by power law.

In previous, simpler treatments of propagating fluctuations, the predicted power spectrum has always consisted of flat top noise between a low and high frequency  break if the total variability power is assumed to vary smoothly with radius. Observed power spectra, on the other hand, routinely show a series of discrete humps, that have been modeled by injecting extra variability at characteristic radii \citep{2016MNRAS.462.4078R,2017arXiv170407705R,2017MNRAS.469.2011R}. Here, however, we show our more sophisticated treatment of the problem naturally leads to discrete humps in the predicted power spectrum, even for input variability varying smoothly with radius. 
{\cite{2017MNRAS.469.2011R} recently showed that observations of XTE J1550-564 pose problems for the propagating fluctuations model, since the power spectra in two energy bands are almost identical, but there are still phase lags between the two bands. This is very challenging to explain, because the lags are assumed to result from the different bands having different radial emissivity laws, which inevitably leads to different power spectral shapes. This is still the case in our more sophisticated treatment of the model. 
{The unique properties of XTE J1550-564 were reported earlier by \citealt{2016A&A...591A..77S} who reported on the evidence for non-linear dynamics in the lightcurves of the source.}
However, the more nuanced predictions explored here leave more scope for the XTE J1550-564 data to be explained. In particular, if the property seen in the observations of \cite{2017MNRAS.469.2011R} is fairly rare, it is reasonable to fit it with rather ``fine tuned" parameters. We also note that the same observation is a problem for all lag models in the literature (e.g. \citealt{2015A&A...584A.109R}). }

The model in the linear regime, i.e., when the initial fluctuations are not affected by mass accretion rate variability, reproduces the observed linear relation between rms and luminosity \citep{2005MNRAS.359..345U}. However, it does not exclude the possibility of deviation from the linear relation. A deviation can be caused by nonlinear interaction between the fluctuations of mass accretion rate. 
We predict deviation from the linear rms-flux relation for very high levels of rms variability. Discovery of such a property in observational data would allow us to break degeneracies associated with propagating fluctuation models.

The variability at smaller radii usually lags variability at larger radii, but we show that at high frequency the situation might be opposite and variability at larger radii lags variability from the inner parts of accretion disc. It leads to the possibility of negative time-lags in BH XRBs: the variability in soft X-rays lags variability in hard X-rays at sufficiently high Fourier frequencies. The typical value of the corresponding negative phase lags is $\sim 0.05\,{\rm rad}$, which is close to the lags observed in AGNs \citep{2010MNRAS.401.2419Z,2013ApJ...764L...9C,2014MNRAS.439.1548A} and stellar-mass black holes \citep{2011MNRAS.414L..60U,2015ApJ...814...50D}. The negative lags are commonly explained by Comptonization time delays \citep{1997ApJ...480..735K} or by Compton-scattering reverberation from material located at some distance from the primary X-ray source \citep{2009Natur.459..540F,2012ApJ...760...73L}. 
Thus, we conclude that viscous diffusion can potentially contribute significantly to the observed lags of soft X-ray radiation.

\section*{Acknowledgments}

This research was supported by the Netherlands Organization for Scientific Research (AAM) and Veni grant (AI). Partial support comes from the EU COST Action MP1304 ``NewCompStar". The authors are grateful to Galina Lipunova and Andrey Semena for useful discussions.


\begin{thebibliography}{}

\bibitem[\protect\citeauthoryear{{Alston}, {Done} \& {Vaughan}}{{Alston}
  et~al.}{2014}]{2014MNRAS.439.1548A}
{Alston} W.~N.,  {Done} C.,    {Vaughan} S.,  2014, \mnras, 439, 1548

\bibitem[\protect\citeauthoryear{{Balbus} \& {Hawley}}{{Balbus} \&
  {Hawley}}{1991}]{1991ApJ...376..214B}
{Balbus} S.~A.,  {Hawley} J.~F.,  1991, \apj, 376, 214

\bibitem[\protect\citeauthoryear{{Brandenburg}, {Nordlund}, {Stein} \&
  {Torkelsson}}{{Brandenburg} et~al.}{1995}]{1995ApJ...446..741B}
{Brandenburg} A.,  {Nordlund} A.,  {Stein} R.~F.,    {Torkelsson} U.,  1995,
  \apj, 446, 741

\bibitem[\protect\citeauthoryear{{Cackett}, {Fabian}, {Zogbhi}, {Kara},
  {Reynolds} \& {Uttley}}{{Cackett} et~al.}{2013}]{2013ApJ...764L...9C}
{Cackett} E.~M.,  {Fabian} A.~C.,  {Zogbhi} A.,  {Kara} E.,  {Reynolds} C.,
  {Uttley} P.,  2013, \apjl, 764, L9

\bibitem[\protect\citeauthoryear{{Churazov}, {Gilfanov} \&
  {Revnivtsev}}{{Churazov} et~al.}{2001}]{2001MNRAS.321..759C}
{Churazov} E.,  {Gilfanov} M.,    {Revnivtsev} M.,  2001, \mnras, 321, 759

\bibitem[\protect\citeauthoryear{{Cowperthwaite} \& {Reynolds}}{{Cowperthwaite}
  \& {Reynolds}}{2014}]{2014ApJ...791..126C}
{Cowperthwaite} P.~S.,  {Reynolds} C.~S.,  2014, \apj, 791, 126

\bibitem[\protect\citeauthoryear{{De Marco}, {Ponti}, {Cappi}, {Dadina},
  {Uttley}, {Cackett}, {Fabian} \& {Miniutti}}{{De Marco}
  et~al.}{2013}]{2013MNRAS.431.2441D}
{De Marco} B.,  {Ponti} G.,  {Cappi} M.,  {Dadina} M.,  {Uttley} P.,  {Cackett}
  E.~M.,  {Fabian} A.~C.,    {Miniutti} G.,  2013, \mnras, 431, 2441

\bibitem[\protect\citeauthoryear{{De Marco}, {Ponti}, {Mu{\~n}oz-Darias} \&
  {Nandra}}{{De Marco} et~al.}{2015}]{2015ApJ...814...50D}
{De Marco} B.,  {Ponti} G.,  {Mu{\~n}oz-Darias} T.,    {Nandra} K.,  2015,
  \apj, 814, 50

\bibitem[\protect\citeauthoryear{{Emmanoulopoulos}, {McHardy} \&
  {Papadakis}}{{Emmanoulopoulos} et~al.}{2011}]{2011MNRAS.416L..94E}
{Emmanoulopoulos} D.,  {McHardy} I.~M.,    {Papadakis} I.~E.,  2011, \mnras,
  416, L94

\bibitem[\protect\citeauthoryear{{Fabian}, {Zoghbi}, {Ross}, {Uttley}, {Gallo},
  {Brandt}, {Blustin}, {Boller}, {Caballero-Garcia}, {Larsson}, {Miller},
  {Miniutti}, {Ponti}, {Reis}, {Reynolds}, {Tanaka} \& {Young}}{{Fabian}
  et~al.}{2009}]{2009Natur.459..540F}
{Fabian} A.~C. et al.,  2009, \nat, 459, 540

\bibitem[\protect\citeauthoryear{{Frank}, {King} \& {Raine}}{{Frank}
  et~al.}{2002}]{2002apa..book.....F}
{Frank} J.,  {King} A.,    {Raine} D.~J.,  2002, {Accretion Power in
  Astrophysics: Third Edition}

\bibitem[\protect\citeauthoryear{{Gaskell}}{{Gaskell}}{2004}]{2004ApJ...612L..21G}
{Gaskell} C.~M.,  2004, \apjl, 612, L21

\bibitem[\protect\citeauthoryear{{Gilfanov} \& {Arefiev}}{{Gilfanov} \&
  {Arefiev}}{2005}]{2005astro.ph..1215G}
{Gilfanov} M.,  {Arefiev} V.,  2005, astro-ph/0501215

\bibitem[\protect\citeauthoryear{{Hawley}, {Gammie} \& {Balbus}}{{Hawley}
  et~al.}{1995}]{1995ApJ...440..742H}
{Hawley} J.~F.,  {Gammie} C.~F.,    {Balbus} S.~A.,  1995, \apj, 440, 742

\bibitem[\protect\citeauthoryear{{Hogg} \& {Reynolds}}{{Hogg} \&
  {Reynolds}}{2016}]{2015arXiv151205350H}
{Hogg} J.~D.,  {Reynolds} C.~S.,  2016, \apj, 826, 40

\bibitem[\protect\citeauthoryear{{Ingram}}{{Ingram}}{2015}]{2015ebha.confE...8I}
{Ingram} A.,  2015, in The Extremes of Black Hole Accretion {Review of X-ray
  variability in black hole binaries}.
p.~8

\bibitem[\protect\citeauthoryear{{Ingram} \& {Done}}{{Ingram} \&
  {Done}}{2011}]{2011MNRAS.415.2323I}
{Ingram} A.,  {Done} C.,  2011, \mnras, 415, 2323

\bibitem[\protect\citeauthoryear{{Ingram} \& {van der Klis}}{{Ingram} \& {van
  der Klis}}{2013}]{2013MNRAS.434.1476I}
{Ingram} A.,  {van der Klis} M.,  2013, \mnras, 434, 1476

\bibitem[\protect\citeauthoryear{{Inogamov} \& {Sunyaev}}{{Inogamov} \&
  {Sunyaev}}{2010}]{2010AstL...36..848I}
{Inogamov} N.~A.,  {Sunyaev} R.~A.,  2010, Astronomy Letters, 36, 848

\bibitem[\protect\citeauthoryear{{Janiuk} \& {Misra}}{{Janiuk} \&
  {Misra}}{2012}]{2012A&A...540A.114J}
{Janiuk} A.,  {Misra} R.,  2012, \aap, 540, A114

\bibitem[\protect\citeauthoryear{{Kazanas}, {Hua} \& {Titarchuk}}{{Kazanas}
  et~al.}{1997}]{1997ApJ...480..735K}
{Kazanas} D.,  {Hua} X.-M.,    {Titarchuk} L.,  1997, \apj, 480, 735

\bibitem[\protect\citeauthoryear{{King}, {Pringle}, {West} \& {Livio}}{{King}
  et~al.}{2004}]{2004MNRAS.348..111K}
{King} A.~R.,  {Pringle} J.~E.,  {West} R.~G.,    {Livio} M.,  2004, \mnras,
  348, 111

\bibitem[\protect\citeauthoryear{{King} \& {Ritter}}{{King} \&
  {Ritter}}{1998}]{1998MNRAS.293L..42K}
{King} A.~R.,  {Ritter} H.,  1998, \mnras, 293, L42

\bibitem[\protect\citeauthoryear{{Kotov}, {Churazov} \& {Gilfanov}}{{Kotov}
  et~al.}{2001}]{2001MNRAS.327..799K}
{Kotov} O.,  {Churazov} E.,    {Gilfanov} M.,  2001, \mnras, 327, 799

\bibitem[\protect\citeauthoryear{{Lasota}}{{Lasota}}{2001}]{2001NewAR..45..449L}
{Lasota} J.-P.,  2001, \nar, 45, 449

\bibitem[\protect\citeauthoryear{{Legg}, {Miller}, {Turner}, {Giustini},
  {Reeves} \& {Kraemer}}{{Legg} et~al.}{2012}]{2012ApJ...760...73L}
{Legg} E.,  {Miller} L.,  {Turner} T.~J.,  {Giustini} M.,  {Reeves} J.~N.,
  {Kraemer} S.~B.,  2012, \apj, 760, 73

\bibitem[\protect\citeauthoryear{{Lightman} \& {Eardley}}{{Lightman} \&
  {Eardley}}{1974}]{1974ApJ...187L...1L}
{Lightman} A.~P.,  {Eardley} D.~M.,  1974, \apjl, 187, L1

\bibitem[\protect\citeauthoryear{{Lipunova}}{{Lipunova}}{2015}]{2015ApJ...804...87L}
{Lipunova} G.~V.,  2015, \apj, 804, 87

\bibitem[\protect\citeauthoryear{{Livio}, {Pringle} \& {King}}{{Livio}
  et~al.}{2003}]{2003ApJ...593..184L}
{Livio} M.,  {Pringle} J.~E.,    {King} A.~R.,  2003, \apj, 593, 184

\bibitem[\protect\citeauthoryear{{L\"ust}}{{L\"ust}}{1952}]{1952Lust}
{L\"ust} R.~Z.,  1952, Zeitschrift Naturforschung Teil A, 7, 87

\bibitem[\protect\citeauthoryear{{Lynden-Bell} \& {Pringle}}{{Lynden-Bell} \&
  {Pringle}}{1974}]{1974MNRAS.168..603L}
{Lynden-Bell} D.,  {Pringle} J.~E.,  1974, \mnras, 168, 603

\bibitem[\protect\citeauthoryear{{Lyubarskii}}{{Lyubarskii}}{1997}]{1997MNRAS.292..679L}
{Lyubarskii} Y.~E.,  1997, \mnras, 292, 679

\bibitem[\protect\citeauthoryear{{Maccarone}, {Coppi} \&
  {Poutanen}}{{Maccarone} et~al.}{2000}]{2000ApJ...537L.107M}
{Maccarone} T.~J.,  {Coppi} P.~S.,    {Poutanen} J.,  2000, \apjl, 537, L107

\bibitem[\protect\citeauthoryear{{Malanchev} \& {Shakura}}{{Malanchev} \&
  {Shakura}}{2015}]{2015arXiv151102356M}
{Malanchev} K.~L.,  {Shakura} N.~I.,  2015, Astronomy Letters, 41, 797

\bibitem[\protect\citeauthoryear{{McHardy}, {Ar{\'e}valo}, {Uttley},
  {Papadakis}, {Summons}, {Brinkmann} \& {Page}}{{McHardy}
  et~al.}{2007}]{2007MNRAS.382..985M}
{McHardy} I.~M. et al.,  2007, \mnras, 382, 985

\bibitem[\protect\citeauthoryear{{McHardy}, {Papadakis}, {Uttley}, {Page} \&
  {Mason}}{{McHardy} et~al.}{2004}]{2004MNRAS.348..783M}
{McHardy} I.~M.,  {Papadakis} I.~E.,  {Uttley} P.,  {Page} M.~J.,    {Mason}
  K.~O.,  2004, \mnras, 348, 783

\bibitem[\protect\citeauthoryear{{Mushtukov}, {Suleimanov}, {Tsygankov} \&
  {Ingram}}{{Mushtukov} et~al.}{2017}]{2017MNRAS.tmp..143M}
{Mushtukov} A.~A.,  {Suleimanov} V.~F.,  {Tsygankov} S.~S.,    {Ingram} A.,
  2017, \mnras, 467, 1202

\bibitem[\protect\citeauthoryear{{Negoro} \& {Mineshige}}{{Negoro} \&
  {Mineshige}}{2002}]{2002PASJ...54L..69N}
{Negoro} H.,  {Mineshige} S.,  2002, \pasj, 54, L69

\bibitem[\protect\citeauthoryear{{Nolan}, {Gruber}, {Matteson}, {Peterson},
  {Rothschild}, {Doty}, {Levine}, {Lewin} \& {Primini}}{{Nolan}
  et~al.}{1981}]{1981ApJ...246..494N}
{Nolan} P.~L.,  {Gruber} D.~E.,  {Matteson} J.~L.,  {Peterson} L.~E.,
  {Rothschild} R.~E.,  {Doty} J.~P.,  {Levine} A.~M.,  {Lewin} W.~H.~G.,
  {Primini} F.~A.,  1981, \apj, 246, 494

\bibitem[\protect\citeauthoryear{{Nowak}, {Vaughan}, {Wilms}, {Dove} \&
  {Begelman}}{{Nowak} et~al.}{1999}]{1999ApJ...510..874N}
{Nowak} M.~A.,  {Vaughan} B.~A.,  {Wilms} J.,  {Dove} J.~B.,    {Begelman}
  M.~C.,  1999, \apj, 510, 874

\bibitem[\protect\citeauthoryear{{Poutanen}}{{Poutanen}}{2001}]{2001AIPC..599..310P}
{Poutanen} J.,  2001, X-ray Astronomy: Stellar Endpoints, AGN, and the Diffuse
  X-ray Background, 599, 310

\bibitem[\protect\citeauthoryear{{Priedhorsky}, {Garmire}, {Rothschild},
  {Boldt}, {Serlemitsos} \& {Holt}}{{Priedhorsky}
  et~al.}{1979}]{1979ApJ...233..350P}
{Priedhorsky} W.,  {Garmire} G.~P.,  {Rothschild} R.,  {Boldt} E.,
  {Serlemitsos} P.,    {Holt} S.,  1979, \apj, 233, 350

\bibitem[\protect\citeauthoryear{{Pringle}}{{Pringle}}{1991}]{1991MNRAS.248..754P}
{Pringle} J.~E.,  1991, \mnras, 248, 754

\bibitem[\protect\citeauthoryear{{Rapisarda}, {Ingram}, {Kalamkar} \& {van der
  Klis}}{{Rapisarda} et~al.}{2016}]{2016MNRAS.462.4078R}
{Rapisarda} S.,  {Ingram} A.,  {Kalamkar} M.,    {van der Klis} M.,  2016,
  \mnras, 462, 4078

\bibitem[\protect\citeauthoryear{{Rapisarda}, {Ingram} \& {van der
  Klis}}{{Rapisarda} et~al.}{2017a}]{2017MNRAS.469.2011R}
{Rapisarda} S.,  {Ingram} A.,    {van der Klis} M.,  2017a, \mnras, 469, 2011

\bibitem[\protect\citeauthoryear{{Rapisarda}, {Ingram} \& {van der
  Klis}}{{Rapisarda} et~al.}{2017b}]{2017arXiv170407705R}
{Rapisarda} S.,  {Ingram} A.,    {van der Klis} M.,  2017b, arXiv:1704.07705

\bibitem[\protect\citeauthoryear{{Reig} \& {Kylafis}}{{Reig} \&
  {Kylafis}}{2015}]{2015A&A...584A.109R}
{Reig} P.,  {Kylafis} N.~D.,  2015, \aap, 584, A109

\bibitem[\protect\citeauthoryear{{Revnivtsev}, {Churazov}, {Postnov} \&
  {Tsygankov}}{{Revnivtsev} et~al.}{2009}]{2009A&A...507.1211R}
{Revnivtsev} M.,  {Churazov} E.,  {Postnov} K.,    {Tsygankov} S.,  2009, \aap,
  507, 1211

\bibitem[\protect\citeauthoryear{{Revnivtsev}, {Gilfanov} \&
  {Churazov}}{{Revnivtsev} et~al.}{2000}]{2000A&A...363.1013R}
{Revnivtsev} M.,  {Gilfanov} M.,    {Churazov} E.,  2000, \aap, 363, 1013

\bibitem[\protect\citeauthoryear{{Sakimoto} \& {Coroniti}}{{Sakimoto} \&
  {Coroniti}}{1989}]{1989ApJ...342...49S}
{Sakimoto} P.~J.,  {Coroniti} F.~V.,  1989, \apj, 342, 49

\bibitem[\protect\citeauthoryear{{Scaringi}, {K{\"o}rding}, {Groot}, {Uttley},
  {Marsh}, {Knigge}, {Maccarone} \& {Dhillon}}{{Scaringi}
  et~al.}{2013}]{2013MNRAS.431.2535S}
{Scaringi} S. et al.,  2013, \mnras, 431, 2535

\bibitem[\protect\citeauthoryear{{Scaringi}, {Manara}, {Barenfeld}, {Groot},
  {Isella}, {Kenworthy}, {Knigge}, {Maccarone}, {Ricci} \&
  {Ansdell}}{{Scaringi} et~al.}{2016}]{2016MNRAS.463.2265S}
{Scaringi} S. et al.,  2016, \mnras, 463, 2265

\bibitem[\protect\citeauthoryear{{Shakura}}{{Shakura}}{1972}]{1972AZh....49..921S}
{Shakura} N.~I.,  1972, \azh, 49, 921

\bibitem[\protect\citeauthoryear{{Shakura} \& {Sunyaev}}{{Shakura} \&
  {Sunyaev}}{1973}]{1973A&A....24..337S}
{Shakura} N.~I.,  {Sunyaev} R.~A.,  1973, \aap, 24, 337

\bibitem[\protect\citeauthoryear{{Sukov{\'a}} \& {Janiuk}}{{Sukov{\'a}} \&
  {Janiuk}}{2016}]{2016A&A...591A..77S}
{Sukov{\'a}} P.,  {Janiuk} A.,  2016, \aap, 591, A77

\bibitem[\protect\citeauthoryear{{Suleimanov}, {Lipunova} \&
  {Shakura}}{{Suleimanov} et~al.}{2007}]{2007ARep...51..549S}
{Suleimanov} V.~F.,  {Lipunova} G.~V.,    {Shakura} N.~I.,  2007, Astronomy
  Reports, 51, 549

\bibitem[\protect\citeauthoryear{{Tanaka}}{{Tanaka}}{2011}]{2011MNRAS.410.1007T}
{Tanaka} T.,  2011, \mnras, 410, 1007

\bibitem[\protect\citeauthoryear{{Tsygankov}, {Mushtukov}, {Suleimanov},
  {Doroshenko}, {Abolmasov}, {Lutovinov} \& {Poutanen}}{{Tsygankov}
  et~al.}{2017}]{2017arXiv170304528T}
{Tsygankov} S.~S. et al.,  2017,
  arXiv:1703.04528

\bibitem[\protect\citeauthoryear{{Uttley}}{{Uttley}}{2004}]{2004MNRAS.347L..61U}
{Uttley} P.,  2004, \mnras, 347, L61

\bibitem[\protect\citeauthoryear{{Uttley}, {Cackett}, {Fabian}, {Kara} \&
  {Wilkins}}{{Uttley} et~al.}{2014}]{2014A&ARv..22...72U}
{Uttley} P.,  {Cackett} E.~M.,  {Fabian} A.~C.,  {Kara} E.,    {Wilkins} D.~R.,
   2014, \aapr, 22, 72

\bibitem[\protect\citeauthoryear{{Uttley} \& {McHardy}}{{Uttley} \&
  {McHardy}}{2001}]{2001MNRAS.323L..26U}
{Uttley} P.,  {McHardy} I.~M.,  2001, \mnras, 323, L26

\bibitem[\protect\citeauthoryear{{Uttley}, {McHardy} \& {Vaughan}}{{Uttley}
  et~al.}{2005}]{2005MNRAS.359..345U}
{Uttley} P.,  {McHardy} I.~M.,    {Vaughan} S.,  2005, \mnras, 359, 345

\bibitem[\protect\citeauthoryear{{Uttley}, {Wilkinson}, {Cassatella}, {Wilms},
  {Pottschmidt}, {Hanke} \& {B{\"o}ck}}{{Uttley}
  et~al.}{2011}]{2011MNRAS.414L..60U}
{Uttley} P. et al.,  2011, \mnras, 414, L60

\bibitem[\protect\citeauthoryear{{Walton}, {Zoghbi}, {Cackett}, {Uttley},
  {Harrison}, {Fabian}, {Kara}, {Miller}, {Reis} \& {Reynolds}}{{Walton}
  et~al.}{2013}]{2013ApJ...777L..23W}
{Walton} D.~J. et al.,  2013, \apjl, 777, L23

\bibitem[\protect\citeauthoryear{{Wood}, {Titarchuk}, {Ray}, {Wolff},
  {Lovellette} \& {Bandyopadhyay}}{{Wood} et~al.}{2001}]{2001ApJ...563..246W}
{Wood} K.~S.,  {Titarchuk} L.,  {Ray} P.~S.,  {Wolff} M.~T.,  {Lovellette}
  M.~N.,    {Bandyopadhyay} R.~M.,  2001, \apj, 563, 246

\bibitem[\protect\citeauthoryear{{Zdziarski}, {Kawabata} \&
  {Mineshige}}{{Zdziarski} et~al.}{2009}]{2009MNRAS.399.1633Z}
{Zdziarski} A.~A.,  {Kawabata} R.,    {Mineshige} S.,  2009, \mnras, 399, 1633

\bibitem[\protect\citeauthoryear{{Zoghbi}, {Fabian}, {Uttley}, {Miniutti},
  {Gallo}, {Reynolds}, {Miller} \& {Ponti}}{{Zoghbi}
  et~al.}{2010}]{2010MNRAS.401.2419Z}
{Zoghbi} A. et al.,  2010, \mnras, 401, 2419

\end{thebibliography}

{

}

\appendix

\section{Mass accretion rate at the outer region}
\label{App:Mdot}

Green function for the mass accretion rate are given by equation (\ref{eq:Gf_MdotLB}) turns to zero only in case when the difference in braces turns to zero. Therefore, one can find the necessary conditions solving the equation 
\be 
I_{l-1}\left(\psi \right)-I_{l}\left( \psi \right)\left(\frac{R}{R'}\right)^{1-n/2}=0. 
\ee
In particular case of $n=1$ the equation takes form 
\be
\cosh\psi-\sinh\psi \left(\frac{R}{R'}\right)^{1-n/2}=0 \nonumber
\ee
or
\be
\tanh\psi=(R'/R)^{1/2}. 
\ee
The equation can be solved only if $R>R'$ because $\tanh x<1$. If $\psi$ and $(R'/R)$ are small enough, the latest equation can be approximated as $\psi\approx(R'/R)^{1/2}$. Therefore
\be
\frac{1}{\tau}\left(\frac{R R'}{R^2_{\rm c}}\right)^{1/2}\approx \left(\frac{R'}{R}\right)^{1/2}
\ee
and $\tau\approx R/R_{\rm c}$ which is dimensionless viscous time for the radius $R>R'$ and $n=1$. As a result, we see that the mass accretion rate at $R>R'$ changes its sing and becomes positive at viscous timescale. 

\begin{figure}
\centering 
\includegraphics[width=8.7cm]{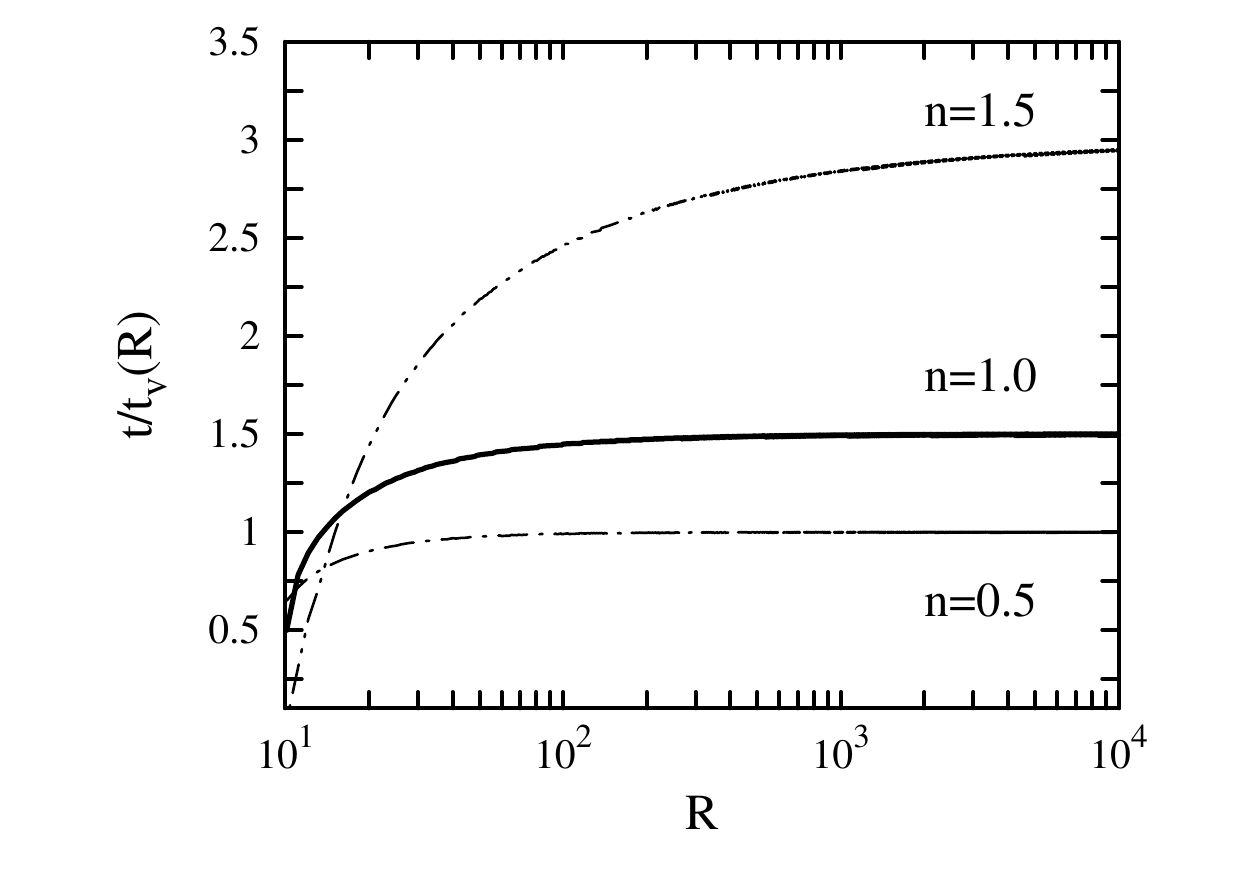} 
\caption{Time scale in units of local viscous time $t_{\rm v}$ at which the mass accretion rate at the outer region of initial perturbation changes its sign and becomes positive. The initial perturbation if located at $R'=10$, $R_{\rm in}=0$. Different curves are given for various parameters $n$. }
\label{pic:backMdot01} 
\end{figure}

The same result can be obtained numerically for various $n\in (0;2)$ (see Fig.\,\ref{pic:backMdot01}).
 
\begin{figure}
\centering 
\includegraphics[width=8.7cm]{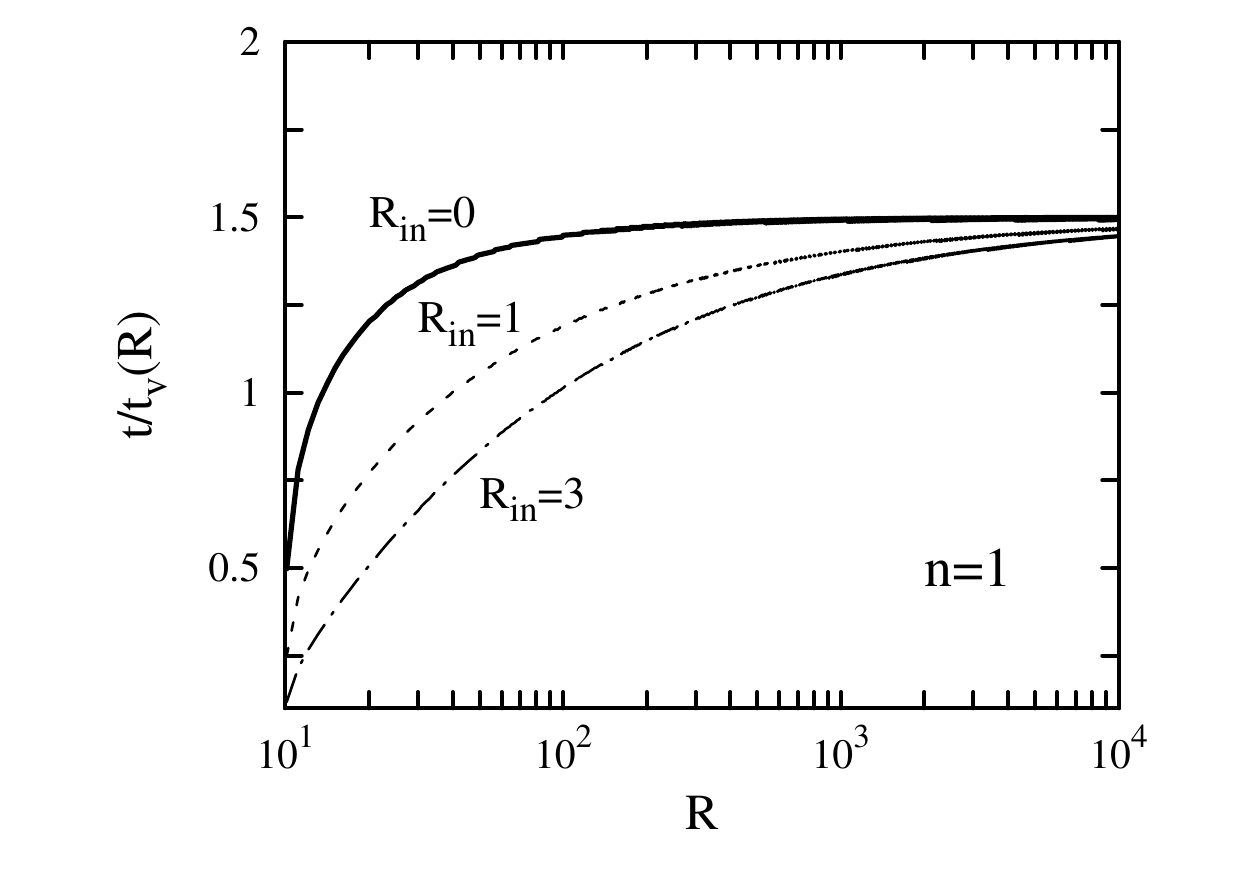} 
\caption{Time scale in units of local viscous time $t_{\rm v}$ at which the mass accretion rate at the outer region of initial perturbation changes its sign and becomes positive (matter distribution is going inwards). The initial perturbation if located at $R'=10$. Solid line is given for the case of accretion disc of $R_{in}=0$, while the dashed and dashed-dotted lines correspond to the disc of $R_{\rm in}=1$ and $R_{\rm in}=3$ respectively. The mass accretion rate changes its sign faster in case of accretion disc with non-zero inner radius. At $R\gg R'$ the time scale at which the mass accretion rate changes its sign $\propto t_{\rm v}$.}
\label{pic:backMdot02} 
\end{figure}

If the inner disc radius $R_{\rm in}>0$ the time scale at which the mass accretion rate change its sing from negative to positive becomes smaller (see Fig.\,\ref{pic:backMdot02}).

\section{The equation in discrete form: The case of arbitrary Green function}
\label{Sec:EqDiscForm2}

The equation for the fluctuations of the mass accretion rate can be obtained in discrete form for the case of an arbitrary Green function in the case of only inwards propagation. Let us use the following designations: 
\beq
\overline{G}_{ij}\equiv \overline{G}_{\dot{M}}(R_i,R_j,f),\quad \overline{a}_{i}\equiv\overline{a}(R_i,f),\nonumber \\ 
\overline{\dot{m}}_i=\overline{\dot{m}}(R_i,f),\quad \xi\equiv \frac{1}{\dot{M}_0}<1.
\eeq
The accretion goes only in one direction - towards the central object. The mass accretion rate fluctuation in  ring $(i+1)$ is
\beq
\label{eq01}
\overline{\dot{m}}_{i+1}=\overline{G}_{(i+1)i}\overline{\dot{m}}_i\Delta R_i \\
+\overline{G}_{(i+1)i}\overline{a}_i\frac{\Delta R_i}{R_i}+\xi \overline{G}_{(i+1)i}\left[\overline{a}_i\frac{\Delta R_i}{R_i}\otimes_f \overline{\dot{m}}_i\right], 
\nonumber
\eeq
where the first term describes perturbations in ring $i$, which were not generated there, but crossed it, the 2nd term describes new perturbations, which were generated on top of average mass accretion rate $\overline{\dot{M}}_0$ and the 3rd term describes perturbations originated in ring $i$ on top of perturbation which originated from outer rings. Perturbations originated in the ring $(i+1)$ are not included in $\overline{\dot{m}}_{i+1}$ by definition. It is assumed that they are originated in a ring, propagate inwards and influence on the inner rings. 


According to (\ref{eq01}):
\be
 \overline{\dot{m}}_{2}=\overline{G}_{21}\overline{\dot{m}}_1\Delta R_1+\overline{G}_{21}\overline{a}_1\frac{\Delta R_1}{R_1}+\xi \overline{G}_{21}\left[\overline{a}_1\frac{\Delta R_1}{R_1}\otimes_f \overline{\dot{m}}_1\right].\nonumber
\ee
If the mass accretion rate perturbations in the 1st ring are caused by processes in it only (there are no perturbations coming from outer rings) then $\overline{\dot{m}}_1=0$ and
\be
 \overline{\dot{m}}_{2}=\overline{G}_{21}\overline{a}_1 \frac{\Delta R_1}{R_1}.\nonumber
\ee
For the 3rd ring we have then
\beq
 \overline{\dot{m}}_{3}&=&\overline{G}_{32}\overline{\dot{m}}_2\Delta R_2+\overline{G}_{32}\overline{a}_2\frac{\Delta R_2}{R_2}+\xi \overline{G}_{32}[\overline{a}_2\frac{\Delta R_2}{R_2}\otimes_f \overline{\dot{m}}_2]  \nonumber \\
          &=&\overline{G}_{32}\overline{G}_{21}\overline{a}_1\frac{\Delta R_1}{R_1}+\overline{G}_{32}\overline{a}_2\frac{\Delta R_2}{R_2} \nonumber \\
      && +\xi \overline{G}_{32}[\overline{a}_2\frac{\Delta R_2}{R_2}\otimes_f \overline{G}_{21}\overline{a}_1\frac{\Delta R_1}{R_1}] \nonumber,
 \eeq
or 
\be\label{eq03}
\overline{\dot{m}}_{3}= \overline{G}_{31}\overline{a}_1\frac{\Delta R_1}{R_1}+\overline{G}_{32}\overline{a}_2\frac{\Delta R_2}{R_2}+\xi \overline{G}_{32}\left[\overline{a}_2\frac{\Delta R_2}{R_2}\otimes_f \overline{G}_{21}\overline{a}_1\frac{\Delta R_1}{R_1}\right].\nonumber
\ee
Finally we get for the innermost radius:
\beq
\label{eq:MultDiscr01}
\overline{\dot{m}}_{N}&=&\sum\limits_{i=1}^{N-1}\frac{\Delta R_i}{R_i}\overline{G}_{Ni}\overline{a}_i \nonumber \\
&&+\xi\sum\limits_{j=1}^{N-2}\sum\limits_{i=j+1}^{N-1}\frac{\Delta R_i}{R_i}\frac{\Delta R_j}{R_j}\overline{G}_{Ni}\overline{a}_i\otimes_f \overline{G}_{ij}\overline{a}_j
\nonumber \\
&&+\xi^2\sum\limits_{j=1}^{N-3}\sum\limits_{i=j+1}^{N-2}\sum\limits_{k=i+1}^{N-1}
\frac{\Delta R_i}{R_i}\frac{\Delta R_j}{R_j}\frac{\Delta R_k}{R_k}\nonumber \\
&&\times \overline{G}_{Nk}\overline{a}_k\otimes_f \overline{G}_{ki}\overline{a}_i\otimes_f \overline{G}_{ij}\overline{a}_j+(...),
\eeq
where the first set of terms describes the linear contribution of initial fluctuations and further sets of terms describe corrections, which arise from nonlinear behavior of fluctuations.

\section{PDS of the mass accretion rate}
\label{App:PSD}

Stationary stochastic process $A(t)$ is equally described by its power density spectrum (PDS) $S_{A}(f)$ and its auto-correlation function 
\be
K_{A}(\tau)=\lim_{T\longrightarrow \infty}\frac{1}{T}\int\limits_{0}^{T}\d tA(t+\tau)A(t)=\left\langle A(t+\tau)A(t)\right\rangle _t,
\ee
where $\langle...\rangle _t$ denotes averaging over time variable. 
PDS and auto-correlation function are connected by the Fourier transformation:
\be\label{eq:PSD_ACF}
S_{A}(f)=\int\limits_{-\infty}^{+\infty}\d\tau\,K_{A}(\tau)e^{-2\pi i f\tau}.
\ee

Let us find PDS for the mass accretion rate, which is defined by the linear part of the equation (\ref{eq:mdot_tdomain}):
$$
\dot{m}(R,t)=\int\limits_{R_{\rm in}}^{R_{\rm out}}\frac{\d R'}{R'}\,G_{\dot{M}}(R,R',t)\otimes_t a(R',t).
$$
The corresponding auto-correlation function is
\beq
K_{\dot{m}}(R,\tau)&=&\langle\dot{m}(R,t+\tau)\dot{m}(R,t)\rangle_{t} \\
&=&\left\langle \int\limits_{R_{\rm in}}^{R_{\rm out}}\frac{\d R'}{R'}
\left[G_{\dot{M}}(R,R',t+\tau)\otimes_t a(R',t+\tau) \right] \right. \nonumber \\
& & \left.\times\int\limits_{R_{\rm in}}^{R_{\rm out}}\frac{\d R''}{R''}
\left[G_{\dot{M}}(R,R'',t)\otimes_t a(R'',t) \right]\right\rangle_{t} \nonumber
\eeq
Integration and finding of arithmetic mean value can be rearranged:
\beq
K_{\dot{m}}(R,\tau)&=& \int\limits_{R_{\rm in}}^{R_{\rm out}}\frac{\d R'}{R'}\int\limits_{R_{\rm in}}^{R_{\rm out}}\frac{\d R''}{R''}
\nonumber \\
&\times &
\left\langle\left[\int\limits_{-\infty}^{+\infty} \d\theta_1 G_{\dot{M}}(R,R',\theta_1) a(R',t+\tau-\theta_1) \right] \right. \nonumber \\
&\times & \left.
\left[\int\limits_{-\infty}^{+\infty}\d\theta_2 G_{\dot{M}}(R,R'',\theta_2) a(R'',t-\theta_2) \right]\right\rangle_{t} \nonumber \\
&=&
\int\limits_{R_{\rm in}}^{R_{\rm out}}\frac{\d R'}{R'}\int\limits_{R_{\rm in}}^{R_{\rm out}}\frac{\d R''}{R''}
\int\limits_{-\infty}^{+\infty} \d\theta_1 \int\limits_{-\infty}^{+\infty} \d\theta_2 \nonumber \\
&\times &
G_{\dot{M}}(R,R',\theta_1)G_{\dot{M}}(R,R'',\theta_2) \nonumber \\
&\times & \left\langle a(R',t+\tau-\theta_1)a(R'',t-\theta_2) \right\rangle_{t}.
\nonumber
\eeq
As a result, the cross-correlation function for $\dot{m}$ can be rewritten:
\beq
K_{\dot{m}}(R,\tau)=
\int\limits_{R_{\rm in}}^{R_{\rm out}}\frac{\d R'}{R'}\int\limits_{R_{\rm in}}^{R_{\rm out}}\frac{\d R''}{R''}
\int\limits_{-\infty}^{+\infty} \d\theta_1 \int\limits_{-\infty}^{+\infty} \d\theta_2 \nonumber \\
\times G_{\dot{M}}(R,R',\theta_1)G_{\dot{M}}(R,R'',\theta_2) K_{a}(R',R'' | \tau-\theta_1+\theta_2),
\eeq
where $K_{a}(R',R'' | \tau-\theta_1+\theta_2)$ is a cross-correlation function of initial perturbations at radii $R'$ and $R''$. If the initial perturbations are not correlated at any $R'\neq R''$, then the auto-correlation function $K_{\dot{m}}(R,\tau)$ turns to zero. In real situation the cross-correlation function of initial perturbations turns to zero only if the radial coordinates are sufficiently distant one from another. At close radii one would expect correlation between initial fluctuations. The exact cross-correlation function will arise from physics of the fluctuations. However, for simplicity we can assume that there is always an \textit{effective} radial scale $\Delta R(R)$ (which can be different at different parts of accretion disc) where the initial perturbations are correlated. Then the cross-correlation function can be approximated by the auto-correlation function of initial perturbations:
$$
K_{a}(R',R'' | \tau)\approx K_{a}(R',\tau),\quad {\rm if}\quad |R'-R''|<\frac{\Delta R(R')}{2}
$$
and
$$
K_{a}(R',R'' | \tau)=0,\quad {\rm if}\quad |R'-R''|>\frac{\Delta R(R')}{2}.
$$
Then if $\Delta R(R')\ll R'$, the auto-correlation function for the mass accretion rate can be represented as follows:
\beq
K_{\dot{m}}(R,\tau)=\int\limits_{R_{\rm in}}^{R_{\rm out}}\frac{\d R'}{(R')^2}\Delta R(R')
\int\limits_{-\infty}^{+\infty} \d\theta_1 \int\limits_{-\infty}^{+\infty} \d\theta_2 \nonumber \\
\times G_{\dot{M}}(R,R',\theta_1)G_{\dot{M}}(R,R',\theta_2) K_{a}(R', \tau-\theta_1+\theta_2).
\eeq
According to (\ref{eq:PSD_ACF}) the power spectrum for the mass accretion rate is
\beq
S_{\dot{m}}(R,f)=\int\limits_{R_{\rm in}}^{R_{\rm out}}\frac{\d R'}{(R')^2}\Delta R(R')
\int\limits_{-\infty}^{+\infty} \d\theta_1 \int\limits_{-\infty}^{+\infty} \d\theta_2
\int\limits_{-\infty}^{+\infty}\d z  \nonumber \\
\times
G_{\dot{M}}(R,R',\theta_1)G_{\dot{M}}(R,R',\theta_2) K_{a}(R',z) e^{-2\pi i f(z+\theta_1-\theta_2)},
\eeq
where $z=\tau-\theta_1+\theta_2$. The integrals over $z$, $\theta_1$ and $\theta_2$ represent Fourier transformation and therefore
\beq
S_{\dot{m}}(R,f)=\int\limits_{R_{\rm in}}^{R_{\rm out}}\frac{\d R'}{(R')^2}\Delta R(R')
\overline{G}_{\dot{M}}(R,R',f)\overline{G}^*_{\dot{M}}(R,R',f)S_{a}(R',f), \nonumber
\eeq
where $\overline{G}_{\dot{M}}(R,R',f)$ is a Green function in frequency domain and $S_{a}(R,f)$ is PDS of initial perturbation of surface density at given radius $R$.
Finally we get:
\beq\label{eq:S01}
S_{\dot{m}}(R,f)=\int\limits_{R_{\rm in}}^{R_{\rm out}}\frac{\d R'}{(R')^2}\Delta R(R')
|\overline{G}_{\dot{M}}(R,R',f)|^2S_{a}(R',f).
\eeq
One can see that the results depends on the radial scale $\Delta R$, where the initial perturbations are correlated. This scale can be different at different radii and contribution of initial perturbations at some regions can be reinforces if the radial scale of correlation there is bigger then in other parts of accretion disc. 

In general case there are two terms in the rhs of equation (\ref{eq:mdot_tdomain}) for perturbations of the mass accretion rate. If the terms are not correlated the resulting PDS is represented by sum of PDS defined by first and second term. Let us consider contribution of the second term in the rhs of (\ref{eq:mdot_tdomain}):
$$
\dot{m}^{\rm II}(R,t)=\frac{1}{\dot{M}_0}\int\limits_{R_{\rm in}}^{R_{\rm out}}\frac{\d R'}{R'}\,G_{\dot{M}}(R,R',t)\otimes_t \left[a(R',t)\dot{m}(R',t)\right].
$$
The auto-correlation function is given by
\beq
&&K^{\rm II}_{\dot{m}}(R,\tau) = \frac{1}{\dot{m}^2}
\int\limits_{R_{\rm in}}^{R_{\rm out}}\frac{\d R'}{R'}\int\limits_{R_{\rm in}}^{R_{\rm out}}\frac{\d R''}{R''}
\int\limits_{-\infty}^{+\infty} \d\theta_1 \int\limits_{-\infty}^{+\infty} \d\theta_2 \nonumber \\
&&\times G_{\dot{M}}(R,R',\theta_1)G_{\dot{M}}(R,R'',\theta_2) K_{a\dot{m}}(R',R'' | \tau-\theta_1+\theta_2) \nonumber
\eeq
If the product $a(R,t)\dot{m}(R,t)$ is correlated with itself only at some radial scale $\Delta R$, the cross-correlation function can be replaced by auto-correlation function.
If $a(t)$ and $\dot{m}(t)$ are not correlated, then the auto-correlation function of their product is represented by
$$
K_{a\dot{m}}(R, \tau)=K_{a}(R, \tau)K_{\dot{m}}(R, \tau)
$$
and we get
\beq
K^{\rm II}_{\dot{m}}(R,\tau) = \frac{1}{\dot{m}^2}
\int\limits_{R_{\rm in}}^{R_{\rm out}}\frac{\d R'}{(R')^2}\Delta R(R')
\int\limits_{-\infty}^{+\infty} \d\theta_1 \int\limits_{-\infty}^{+\infty} \d\theta_2 \nonumber \\
\times G_{\dot{M}}(R,R',\theta_1)G_{\dot{M}}(R,R',\theta_2)\nonumber\\  
\times K_{a}(R', \tau-\theta_1+\theta_2) K^{\rm II}_{\dot{m}}(R', \tau-\theta_1+\theta_2). \nonumber
\eeq
The corresponding PDS is
\beq
S^{\rm II}_{\dot{m}}(R,f)=\frac{1}{\dot{m}^2}
\int\limits_{R_{\rm in}}^{R_{\rm out}}\frac{\d R'}{(R')^2}\Delta R(R')
\int\limits_{-\infty}^{+\infty} \d\theta_1 \int\limits_{-\infty}^{+\infty} \d\theta_2 \int\limits_{-\infty}^{+\infty}\d z \nonumber \\
\times G_{\dot{M}}(R,R',\theta_1)G_{\dot{M}}(R,R',\theta_2)  \nonumber \\
\times
K_{a}(R', z) K^{\rm II}_{\dot{m}}(R', z)e^{-2\pi i f(z+\theta_1-\theta_2)} \nonumber \\
=\frac{1}{\dot{m}^2} \int\limits_{R_{\rm in}}^{R_{\rm out}}\frac{\d R'}{(R')^2}\Delta R(R')
|G_{\dot{M}}(R,R',f)|^2  \nonumber \\
\times
S_{a}(R',f)\otimes_f S^{\rm II}_{\dot{m}}(R',f). \nonumber 
\eeq
Finally, we get equation for PDS of mass accretion rate perturbations given by equation (\ref{eq:mdot_tdomain}):
\beq
S_{\dot{m}}(R,f)=\int\limits_{R_{\rm in}}^{R_{\rm out}}\frac{\d R'}{(R')^2}\Delta R(R')
|\overline{G}(R,R',f)|^2 \nonumber \\
\times 
S_{a}(R',f)\otimes_f \left[\delta(f) + \frac{S_{\dot{m}}(R',f)}{\dot{M}^2_0}\right],
\eeq
where the first and second terms under the integral describe linear and non-linear contribution of initial fluctuations correspondingly and $S_{a}(R,f)$ represents PDS of initial perturbations.

\section{Cross-spectrum of the mass accretion rate}
\label{App:Cross}

The cross-spectrum $C_{x,y}(f)$ for given stochastic processes $x(t)$ and $y(t)$ can be obtained from their cross-covariance 
$$
\gamma_{x,y}(\tau)=\left\langle x(t+\tau)y(t) \right\rangle_t
$$
by using Fourier transform
\be
\label{eq:gamma2C}
C_{x,y}(f)=\int\limits_{-\infty}^{+\infty}\d \tau\,\gamma_{x,y}(\tau) e^{-2\pi i f\tau}.
\ee

Let us denote the cross-covariace of the mass accretion rate variability at radii $R_1$ and $R_2$ by $\gamma_{\dot{m}}(R_1,R_2\,|\,\tau)$. Let us also assume first that the mass accretion rate variability is defined by the first term of equation (\ref{eq:mdot_tdomain}). Then
\beq
\gamma_{\dot{m}}(R_1,R_2\,|\,\tau)&=&\left\langle \dot{m}(R_1,t+\tau)\dot{m}(R_2,t) \right\rangle_t \nonumber \\
&=& \int\limits_{R_{\rm in}}^{R_{\rm out}}\frac{\d R'}{R'}\int\limits_{R_{\rm in}}^{R_{\rm out}}\frac{\d R''}{R''}
\int\limits_{-\infty}^{+\infty} \d\theta_1 \int\limits_{-\infty}^{+\infty} \d\theta_2 \nonumber \\
&&\times  G_{\dot{M}}(R_1,R',\theta_1)G_{\dot{M}}(R_2,R'',\theta_2) \nonumber \\
&&\times \left\langle a(R',\tau+t-\theta_1)a(R'',t-\theta_2) \right\rangle_t \nonumber
\eeq
and therefore
\beq
\gamma_{\dot{m}}(R_1,R_2\,|\,\tau)=
\int\limits_{R_{\rm in}}^{R_{\rm out}}\frac{\d R'}{R'}\int\limits_{R_{\rm in}}^{R_{\rm out}}\frac{\d R''}{R''}
\int\limits_{-\infty}^{+\infty} \d\theta_1 \int\limits_{-\infty}^{+\infty} \d\theta_2 \nonumber \\
\times  G_{\dot{M}}(R_1,R',\theta_1)G_{\dot{M}}(R_2,R'',\theta_2)\nonumber \\
\times \gamma_{a}(R_1,R_2\,|\,\tau-\theta_1+\theta_2),
\eeq
where $\gamma_{a}(R_1,R_2\,|\,t)$ is cross-covariance of initial perturbations at radii $R_1$ and $R_2$. According to (\ref{eq:gamma2C}) the cross-spectrum of mass accretion rate variability at radii $R_1$ and $R_2$ is
\beq
C_{\dot{m}}(R_1,R_2\,|\,f)=\int\limits_{R_{\rm in}}^{R_{\rm out}}\frac{\d R'}{R'}\int\limits_{R_{\rm in}}^{R_{\rm out}}\frac{\d R''}{R''}
\int\limits_{-\infty}^{+\infty} \d\theta_1 \int\limits_{-\infty}^{+\infty} \d\theta_2 \int\limits_{-\infty}^{+\infty} \d z \nonumber \\
\times  G_{\dot{M}}(R_1,R',\theta_1)G_{\dot{M}}(R_2,R'',\theta_2)\nonumber \\
\times \gamma_{a}(R_1,R_2\,|\,z) e^{-2\pi i f(z+\theta_1-\theta_2)}.
\eeq
If the initial perturbations are correlated at radial scale $\Delta R$ the cross-spectrum can be calculated as follows:
\beq
C_{\dot{m}}(R_1,R_2\,|\,f)=\int\limits_{R_{\rm in}}^{R_{\rm out}}\frac{\d R'}{(R')^2}\Delta R(R') \nonumber \\
\times \overline{G}_{\dot{M}}(R_1,R',f)\overline{G}^*_{\dot{M}}(R_2,R',f) S_{a}(R',f),
\eeq
where $S_a(R,f)$ is PDS of initial perturbations at radius $R$ and "$^*$" denotes complex conjugate. 

In more general case, when the local mass accretion rate is defined by non-linear term of equation (\ref{eq:mdot_tdomain}) as well, the cross-spectrum is given by
\beq
C_{\dot{m}}(R_1,R_2\,|\,f)=\int\limits_{R_{\rm in}}^{R_{\rm out}}\frac{\d R'}{(R')^2}\Delta R(R') \nonumber \\
\times \overline{G}_{\dot{M}}(R_1,R',f)\overline{G}^*_{\dot{M}}(R_2,R',f) \nonumber \\
\times
S_{a}(R',f)\otimes_f
\left[\delta(f)+\frac{S_{\dot{m}}(R',f)}{\dot{M}^2_0}\right].\nonumber
\eeq
Now the cross-spectrum depends also on the PDS of local variations of the mass accretion rate $S_{\dot{m}}(R,f)$, which has to be found from non-linear equation (\ref{eq:S_mdot}).

\section{Power spectrum and cross-spectrum of the photon flux}

The power spectrum of X-ray flux at given energy band can be obtained as a Fourier transform of corresponding auto-correlation function.
According to (\ref{eq:Flux}) the X-ray flux auto-correlation function is
\be
K_{\rm h}(\tau)=\left\langle
\int\limits_{R_{\rm in}}^{R_{\rm out}}\d R'\,\frac{h(R')}{R'^2}\dot{m}(R',t)
\int\limits_{R_{\rm in}}^{R_{\rm out}}\d R''\,\frac{h(R'')}{R''^2}\dot{m}(R'',t)
\right\rangle _{t}.
\nonumber
\ee
This relation can be reduced to 
\be
K_{\rm h}(\tau)=
\int\limits_{R_{\rm in}}^{R_{\rm out}}\d R'
\int\limits_{R_{\rm in}}^{R_{\rm out}}\d R''
\,\frac{h(R')}{R'^2} \frac{h(R'')}{R''^2} 
K_{\dot{m}}(R',R''\,|\,\tau).
\ee
Fourier transform gives the flux power spectrum for given energy band:
\be
S_{\rm h}(f)= 
\int\limits_{R_{\rm in}}^{R_{\rm out}}\d R'
\int\limits_{R_{\rm in}}^{R_{\rm out}}\d R''
\,\frac{h(R')}{R'^2} \frac{h(R'')}{R''^2} 
C_{\dot{m}}(R',R''\,|\,f).
\ee
{One can see that $S_{\rm h}(f)\in \mathbb{R}$ because $C_{\dot{m}}(R_1,R_2\,|\,f)=C^*_{\dot{m}}(R_2,R_1\,|\,f)$.
The cross-spectrum between two energy band is obtained if one of the weighting functions is changed (see equation\,(\ref{eq:CrossEnergyBand})).}

\bsp	
\label{lastpage}
\end{document}